\documentclass{article}
\usepackage{neurips_custom}

\usepackage{cite}
\usepackage{amsmath,amssymb,amsfonts}
\usepackage{microtype}

\usepackage{graphicx}
\usepackage{tabularx}
\usepackage{textcomp}
\usepackage{xcolor}
\usepackage{caption}
\usepackage{algorithm}
\usepackage[noend]{algpseudocode}
\usepackage{scalerel}
\usepackage{subcaption}
\usepackage[nolist,nohyperlinks]{acronym}

\usepackage{tikz, circuitikz}
\usetikzlibrary{arrows,decorations.pathmorphing,backgrounds,positioning,fit,petri,quotes}
\usepackage{import}

\usepackage{hyperref}
\hypersetup{hidelinks, colorlinks=false, linktoc=all, linkcolor=black,}

\title{Delay Conditioned Generative Modelling of Resistive Drift in Memristors}

\acknowledgements{This work was supported by the EPSRC under the DTP 2016-2017 (EP/N509486/1), DTP 2018-2019 (EP/R513052/1), SONATA (EP/W035960/1) and FORTE (EP/R024642/1) projects.}

\author{
  Waleed El-Geresy \\
  Imperial College London\\
  \texttt{waleed.el-geresy15@imperial.ac.uk} \\
  \And
  Christos Papavassiliou \\
  Imperial College London \\
  \texttt{c.papavas@imperial.ac.uk} \\
  \And
  Deniz Gündüz \\
  Imperial College London \\
  \texttt{d.gunduz@imperial.ac.uk} \\
}

\begin{document}

\maketitle

\begin{abstract}

    The modelling of memristive devices is an essential part of the development of novel in-memory computing systems. Models are needed to enable the accurate and efficient simulation of memristor device characteristics, for purposes of testing the performance of the devices or the feasibility of their use in future neuromorphic and in-memory computing architectures. The consideration of memristor non-idealities is an essential part of any modelling approach. The nature of the deviation of memristive devices from their initial state, particularly at ambient temperature and in the absence of a stimulating voltage, is of key interest, as it dictates their reliability as information storage media - a property that is of importance for both traditional storage and neuromorphic applications. In this paper, we investigate the use of a generative modelling approach for the simulation of the delay and initial resistance-conditioned resistive drift distribution of memristive devices. We introduce a data normalisation scheme and a novel training technique to enable the generative model to be conditioned on the continuous inputs. The proposed generative modelling approach is suited for use in end-to-end training and device modelling scenarios, including learned data storage applications, due to its simulation efficiency and differentiability.
    
\end{abstract}

\begin{acronym}[]
        \acro{AWGN}{Additive White Gaussian Noise}
        \acro{AI}{artificial intelligence}

        \acro{BN}{Batch Normalisation}
        \acro{BCE}{Binary Cross-Entropy}

        \acro{cGAN}{conditional Generative Adversarial Network}
        \acro{CPU}{Central Processing Unit}

        \acro{DCGAN}{Deep Convolutional Generative Adversarial Network}
        \acro{Deep JSCC}{Deep Joint Source-Channel Coding}

        \acro{GAN}{Generative Adversarial Network}
        \acro{GMSM}{Generalised Metastable Switch Model}
        \acro{GPU}{Graphics Processing Unit}
        \acro{GRU}{Gated Recurrent Unit}
        \acro{GST}{Germanium Antimony Tellurium}

        \acro{JSCC}{Joint Source-Channel Coding}

        \acro{KL Divergence}{Kullback-Leibler Divergence}

        \acro{LSTM}{Long Short-Term Memory}

        \acro{MIM}{Metal-Insulator-Metal}
        \acro{MSE}{Mean Squared Error}

        \acro{PCM}{Phase Change Memory}
        \acro{PSNR}{Peak Signal to Noise Ratio}

        \acro{RNN}{Recurrent Neural Network}
        \acro{RRAM}{Resistive Random Access Memory}

        \acro{SNR}{Signal to Noise Ratio}
        \acro{SGD}{Stochastic Gradient Descent}

        \acro{WAE}{Wasserstein Autoencoder}
\end{acronym}

\acresetall

\section{Introduction}
\label{sec:introduction}
Memristors are a class of passive electronic device that have a semi-volatile resistive state (memory) which can be modified through the application of a driving voltage or current. The memristor was originally hypothesised by Leon Chua \cite{chuaMemristorTheMissingCircuit1971} as the missing fundamental passive circuit element in a ``periodic table'' of passive electronic devices. Later, the definition and scope of the term ``memristor'' expanded to encompass a variety of resistive switching technologies \cite{chuaIfItPinched2014,chuaResistanceSwitchingMemories2011}, and the definition as used nowadays in literature on the subject is more or less interchangeable with the term ``memristive systems'' \cite{chuaMemristiveDevicesSystems1976} - a generalisation of the original definition. Notable types of device that have been described as ``memristive'' include phase change memory devices and thin film metal oxide devices. The possible uses of memristors go beyond information storage and extend to the realm of so-called neuromorphic computing \cite{sungPerspectiveReviewMemristive2018}, where computation and storage occur at the same physical location, and thus, in a distributed and massively parallelised way \cite{meadNeuromorphicElectronicSystems1990}.

Due to their promising potential applications, building useful models of memristive devices has become a subject of interest. One particular problem is that of modelling the noise and stochasticity (the non-deterministic components) of device behaviour.

The problem of building effective models of memristive devices has been previously addressed using both deterministic \cite{strukovMissingMemristorFound2008, pickettSwitchingDynamicsTitanium2009, yangMemristiveSwitchingMechanism2008} and stochastic approaches \cite{molterGeneralizedMetastableSwitch2016, malikAbsorbingMarkovChain2022}. Modelling considerations include the computational complexity, the accuracy, and the explainability (physical realism and grounding) of the model \cite{williamsPhysicsbasedMemristorModels2013}. In order to provide models that are minimally computationally intensive, but also accurate, we must take into account the intended application. This also has a bearing on the range of applications for the model. For example, models may be written in SPICE to allow for simulation alongside existing analogue circuitry \cite{berdanMemristorSPICEModel2014} and more recently, a model intended for event-based neuromorphic computing applications has been proposed \cite{el-geresyEventBasedSimulationStochastic2024}.

In the context of deep learning and \ac{AI}, data-driven approaches to modelling - known as generative modelling approaches - are becoming increasingly popular due to their ability to learn to simulate complex data with minimal assumptions on the nature of the 
underlying distribution. The popularity of generative modelling techniques has been fuelled by recent impressive results in the conditional generation of realistic data in a variety of modalities, such as natural language, images, and video.

In this work, we develop a generative modelling technique for the modelling of one particular kind of memristive non-ideality: resistive drift. Although a variety of non-idealities plague memristive devices \cite{rumseyCapacityConsiderationsData2019, ohImpactResistanceDrift2019, zarconeAnalogCodingEmerging2020}, resistive drift noise is unique in that it is sequential in time; that is, it has temporal correlations. Resistive drift describes the phenomenon of the stochastic decay of the state of a memristive device over time, until an equilibrium resistance is attained and all information about the original storage values is lost \cite{ielminiModelingUniversalSet2011, el-geresyEventBasedSimulationStochastic2024}. The problem of modelling resistive drift deterministically has been previously addressed through methods including modelling that incorporates autonomous state evolution \cite{ carbajalMemristorModelsMachine2015}, and modelling of the resistive drift using a stretched exponential function \cite{abbeyThermalEffectsInitial2022}. Modelling approaches that take into account the stochastic nature of the state transitions have also been proposed \cite{malikAbsorbingMarkovChain2022, el-geresyEventBasedSimulationStochastic2024}, though so far, generative modelling approaches have not been explored.

\subsection{Time Series Generative Modelling}
\label{sec:time_series_generative_models}

The temporal correlations and dependencies present in time series data present a unique challenge that require specialised time series generative modelling techniques. Time series generative models come in a variety of forms. They can be either recurrent, or non-recurrent, and they can be either deterministic or stochastic.

Recurrent architectures make use of stateful neural networks such as \ac{LSTM} cells \cite{hochreiterLongShortTermMemory1997} or \acp{GRU} \cite{choLearningPhraseRepresentations2014}, which use a state variable to capture dependencies between elements of the series over time. Challenges associated with generating time series using \ac{RNN} architectures also include the computational complexity of the model, with many subsequent sampling steps leading to a large number of forward passes, if the recurrent architecture is unrolled. This increases the computational complexity of the generation of a data sample, and also means that the network is prone to vanishing gradient issues during training, or if used as a fixed differentiable model in an end-to-end training setup.

Alternatively, finite-memory time series generation can be achieved through the use of autoregressive architectures \cite{vandenoordWaveNetGenerativeModel2016}, which generate values based on a fixed horizon of past inputs.
Despite the effectiveness of autoregressive models, they are fundamentally deterministic, being functions of past inputs. In addition, while such models can generate a future value, or sometimes sequences, in a single step, they must be recurrently evaluated on previously generated data to generate values beyond this.

Problems in training both types of time series models may arise due to the disconnect between so-called ``closed-loop" training and ``open-loop" evaluation \cite{yoonTimeseriesGenerativeAdversarial2019}, where the contrast between evaluation of the model in a recurrent manner during inference, as opposed to its conditioning on previous ground truth data values during training, may cause problems with distributional mismatch. Several methods exist for addressing this problem, notably the use of scheduled sampling (a form of curriculum learning), involving a gradual transition between open-loop training and closed-loop training \cite{bengioScheduledSamplingSequence2015}. It has however been suggested that techniques such as scheduled sampling are alone insufficient for addressing this problem \cite{yoonTimeseriesGenerativeAdversarial2019}, with the technique also being criticised for memorisation of samples \cite{huszarHowNotTrain2015}.

\subsection{Contributions and Outline}

In this paper, we employ a generative modelling approach to learn a differentiable model of the resistive drift, conditioned on two continuous conditions: an initial resistance value and a delay.
The use of generative modelling to simulate the resistive drift has the advantage of being purely data driven, and allows us to model the drift over time in a computationally efficient manner, creating a differentiable model that is suitable for use in deep learning frameworks. We propose a computationally simple approach to single-shot generation, which additionally reduces problems associated with vanishing gradients which would be present for models involving recurrence.
We condition directly on the delay rather than recurrently evaluating.
As such, our contributions can be summarised as follows:

\begin{enumerate}
    \item We investigate methods for data normalisation in the context of time series data with a continuous, high dynamic range for the model input and output values. Conditioning \acp{GAN} on continuous values has previously been investigated \cite{dingCcGANContinuousConditional2022}, although in the context of resistive drift, the large dynamic range for time and resistance inputs and resistance outputs presents a particular challenge.
    \item We present an approach to conditioning generative models directly on the delay that exploits the temporal structure of time series data to enforce consistency in time series outputs generated at different timescales.
    Traditional time series generative models are fixed in the timescales at which they generate time series data; that is, they generate time series data of fixed intervals, say \(\Delta\). Therefore, to generate a sample at time \(t + k\cdot\Delta\), the model is run \(k\) times in a recursive fashion.
    We introduce a novel adversarial training technique, which we call \textit{delay discrimination}, to enable us to generate time series data at a range of timescales with consistency.
    By conditioning the model on the delay, we avoid the need for many recurrent evaluations of the model, making it computationally efficient. This minimisation or elimination of additional recurrent inference steps also thus alleviates problems caused by the disconnect between closed-loop training and open-loop evaluation.
    Our method also allows the model to generate outputs at arbitrary continuous delay values, rather than being constrained to generate values at integer multiples of the timestep.
    \item We demonstrate the benefits and effectiveness of the proposed model in end-to-end training settings, detailing its use in a learned approach to the problem of quantised data storage and recovery on memristors.
\end{enumerate}

In Section~\ref{sec:dataset}, we describe the process and motivation behind the creation of an example resistive drift dataset, used to train and evaluate our proposed approach. We make use of a general stochastic memristor model for the creation of the dataset \cite{el-geresyEventBasedSimulationStochastic2024}. In Section~\ref{sec:method}, we introduce the proposed \ac{cGAN} training procedure and methods, including the novel approach of delay discrimination, which is used to enable our network to learn to generate outputs conditioned on a wide range of delays. We present the model fitting results and a discussion of the performance of our method in Section~\ref{sec:results}. Finally, in Section~\ref{sec:quantisation_application}, we demonstrate the use of our modelling approach in the problem of the design of an end-to-end optimised, low-error quantisation scheme, demonstrating the benefits of its computational efficiency and efficacy in a realistic problem.

\textbf{Notation} Throughout this paper, we use \(BCE(y, \hat{y}) \triangleq (y - 1)\ln{(1 - \hat{y})} - y\ln{(\hat{y})} \) to represent the \ac{BCE} function. In general, we use lowercase \(r\) to denote resistance values, \(d\) to denote delay values, and \(\bar{y}\) to denote a normalised version of a given variable \(y\). \(N_x(\cdot)\) is used to denote a particular data normalisation transform.

\section{Resistive Drift Dataset}
\label{sec:dataset}
We make use of the titanium dioxide memristor model developed in \cite{el-geresyEventBasedSimulationStochastic2024} to model resistive drift in a simulated, highly stochastic memristive device. We generate a dataset consisting of 5000 drift time series. We choose a set of parameters that result in relatively unstable device characteristics, presenting a relatively extreme and challenging modelling scenario, where decay to an equilibrium resistance point happens at a rapid rate.

\subsection{Parameter Selection and Dataset Creation}

We choose the parameter \(V_a\), which controls the rate of switching, to yield a drift distribution with sufficient variation and stochasticity over the interval of interest of 1000s. We choose the parameters \(N\), \(n_{\text{thresh}}\), \(g_{\text{step}}\), and \(g_{\text{parallel}}\) to allow for a sufficiently fine description of the states over the resistance interval \([0.1k\Omega, 100k\Omega]\): our chosen active switching region. We choose our equilibrium point to be approximately \(500 k\Omega\), corresponding to \(n=100\) according to the chosen parameters and the chosen readout equation, mimicking the one developed for the titanium dioxide memristor in \cite{el-geresyEventBasedSimulationStochastic2024}:
\begin{align}
    R(n) = \frac{1}{g_{\text{diff}}\max{(n, n_{\text{thresh}})} + \frac{1}{R_{\text{high}}}}
    \label{eq:resistance_readout}
\end{align}
We set \(V_{\text{off}}\) accordingly to achieve this, using Equation~\ref{eq:equilibrium_v_offset}, as derived in \cite{el-geresyEventBasedSimulationStochastic2024}. Using a value of \(n_{eq}=100\) along with the chosen value of \(V_a = 0.256\), in order to adjust the stability/speed of convergence, gives a required value of \(V_{\text{off}} = 0.2533\) to achieve the necessary equilibrium state.

\begin{align}
    V_{\text{off}} = \ln{\left(\frac{N-n_{eq}}{n_{eq}}\right) \cdot \frac{k_B T}{q}}
    \label{eq:equilibrium_v_offset} 
\end{align}

For the given \(V_{\text{off}}\), we invert the equation to calculate the error in our equilibrium point: we can see that this yields gives a true equilibrium value of \(n_{eq}=100.38\), corresponding to a resistance of \(R_{eq}=499.076k\Omega\), which is sufficiently close to the desired value of 500\(k\Omega\).

A summary of the physical parameters and constants chosen for generation of the data are given in Table~\ref{tab:resistance_dataset_params}.

\begin{table}[]
\centering
\caption{The parameters used for generation of the simulated resistance drift dataset, with an equilibrium resistance of 500\(k\Omega\). For a full description of the model used to generate the dataset, please see \cite{el-geresyEventBasedSimulationStochastic2024}.}
\label{tab:resistance_dataset_params}
\begin{tabularx}{\linewidth}{|c|X|c|}
\hline
Parameter & Definition & Value \\
\hline
N & The number of metastable switches, dictating the granularity of simulation. & \(1.8\times10^6\) \\
\(n_{\text{thresh}}\) & The threshold for the resistance readout equation. & 0 \\
\(V_a\) & The energy barrier height, dictating absolute switching rate. & 0.256 \\
\(g_{\text{parallel}}\) & The low conductance in the readout equation. & \(1 \times 10^{-6}\) \\
\(g_{\text{step}}\) & The marginal conductance increase in the readout equation. & \(\frac{1 \times 10^{-2}}{1 \times 10^6}\) \\
\(V_{\text{off}}\) & The offset voltage dictating the equilibrium offset. & 0.2532 \\
T & The bath temperature. & 300 \\
\(k_B\) & Boltzmann's constant. & \(1.38064 \times 10^{-23}\) \\
q & The electronic charge. & \(1.602176 \times 10^{-19}\) \\
\hline
\end{tabularx}
\end{table}

We generate 5000 time series with initial points spaced uniformly in the range \([100\Omega, 750k\Omega]\). For each generated series, we sample the data uniformly,
with timestep \(T_{\text{sample}} = 1\) and total series duration \(T_{\text{tot}} = 1000\), resulting in 1001 data points in each series, spaced one second apart (inclusive of points at \(t=0\) and \(t=T_{\text{tot}}\)).

\subsection{Dataset Visualisation and Evaluation}

\begin{figure}
    \centering
    \hspace*{\fill}
    \subfloat[]{\includegraphics[width=0.45\linewidth]{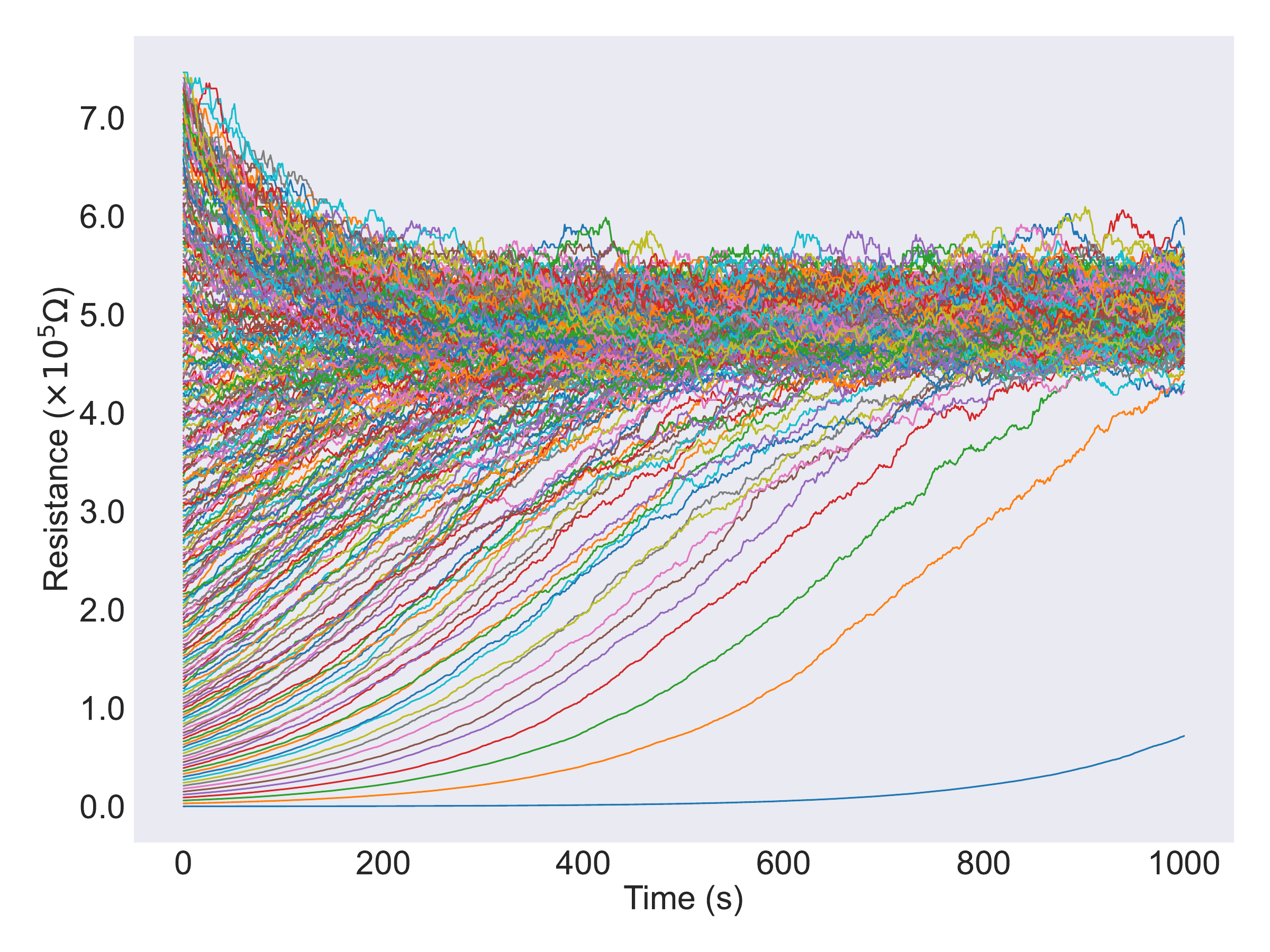}}
    \hfill
    \subfloat[]{\includegraphics[width=0.45\linewidth]{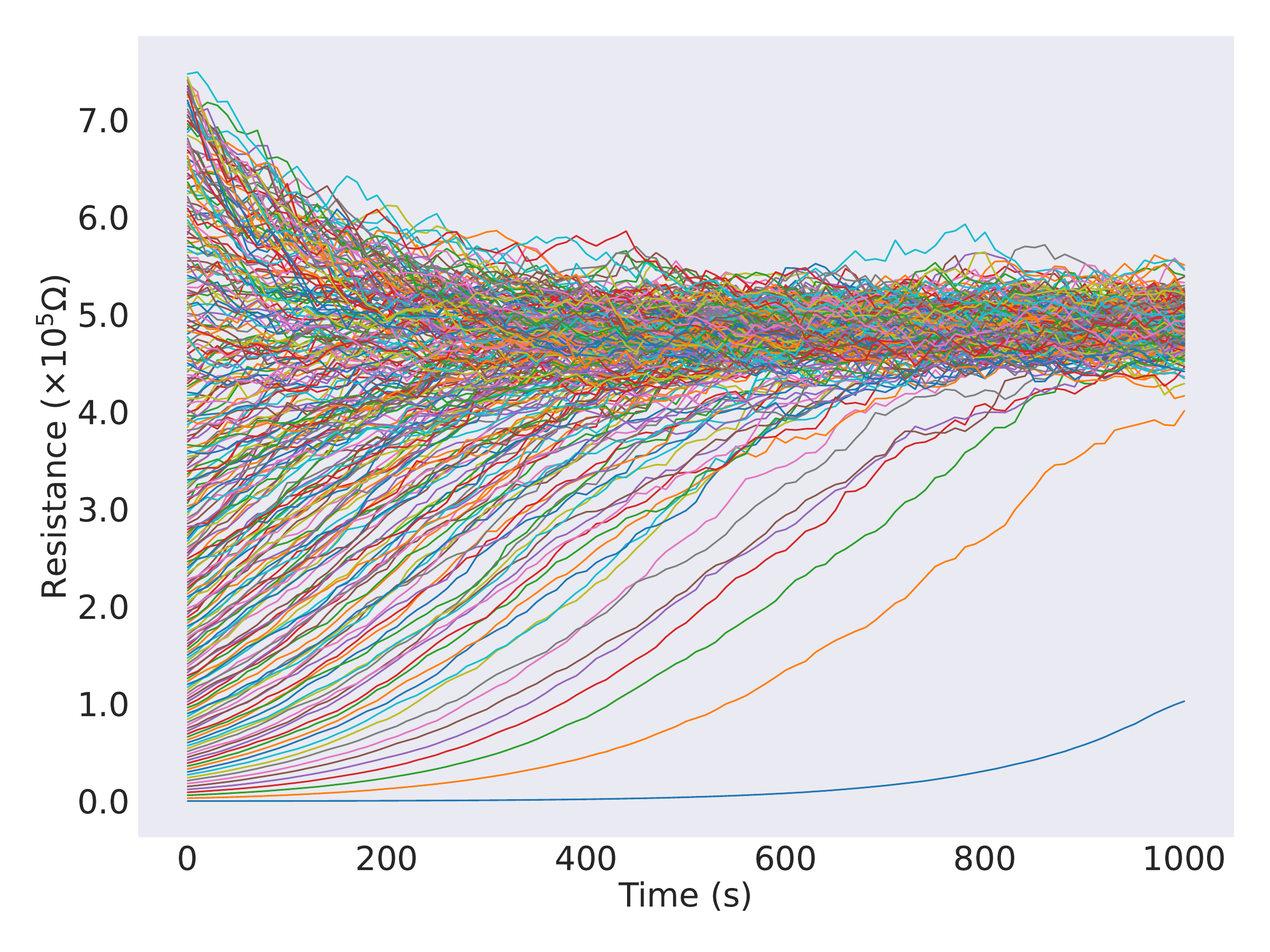}}
    \hspace*{\fill}
    \caption{\textbf{Left}: An illustrative subset of the time series dataset used for training the \ac{cGAN}, generated using the event-based model of \cite{el-geresyEventBasedSimulationStochastic2024}. It clearly illustrates the stochasticity of the resistance values, and their convergence to the equilibrium point of 500\(k\Omega\).
    The variance of the state transitions increases with increasing initial resistance. \textbf{Right}: The proposed modelling approach evaluated on the same initial resistances for a timestep of 10 and a series length of 100 (a total delay of 1000). It is clear that the model is able to approximate the conditional drift distribution.
    }
    \label{fig:drift_data}
    \Description[Time series from the dataset and generated time series.]{A side by side comparison of the time series from the resistive drift dataset and time series generated by the proposed model, showing visually similar behaviour and the same equilibrium point.}
\end{figure}

\begin{figure}
    \centering
    \includegraphics[width=0.8\linewidth]{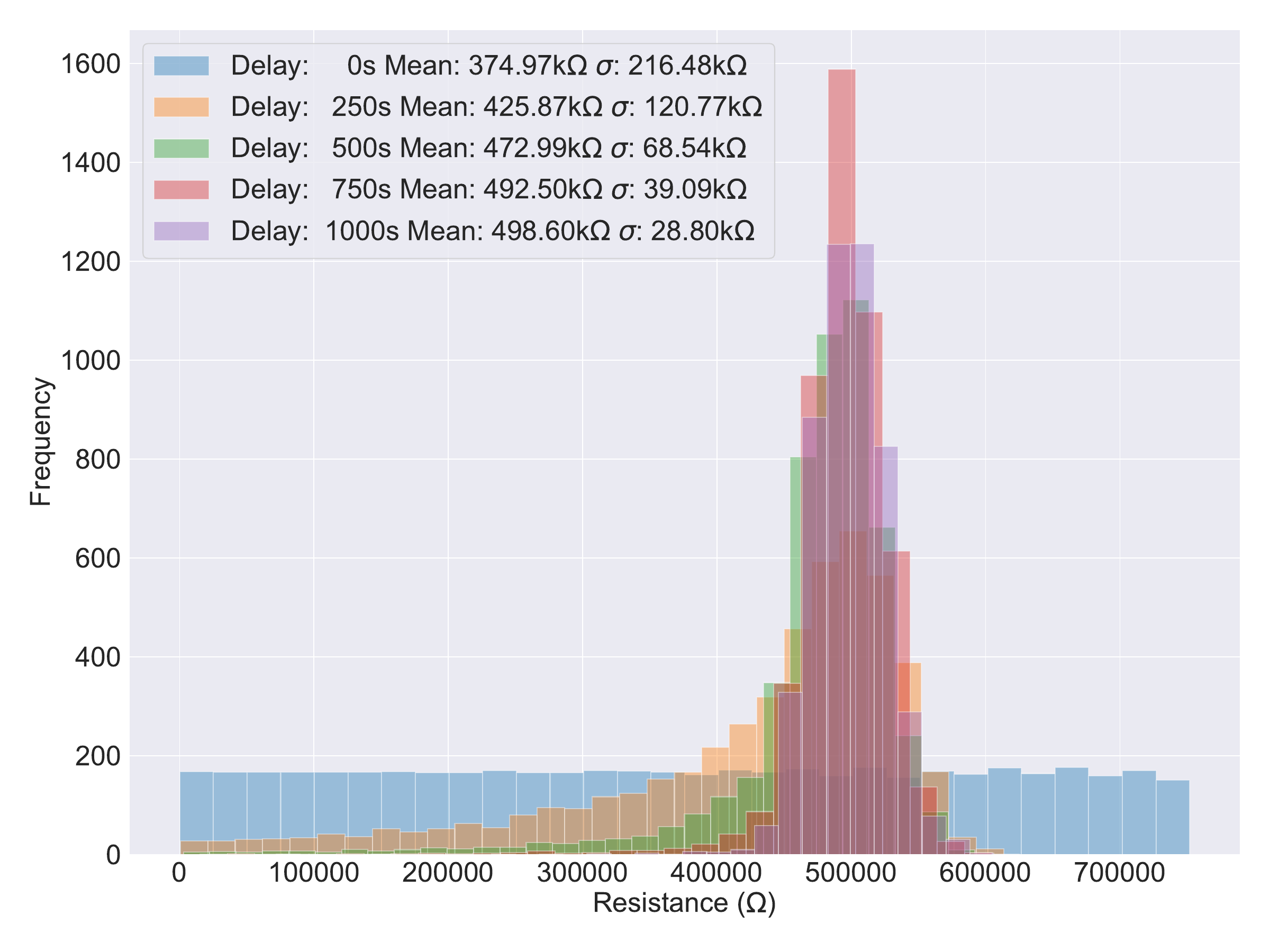}
    \caption{Histogram of final resistance values in the drift dataset at different delays. We see that as the delay increases to 1000s, the distribution becomes centred around the equilibrium resistance value - i.e. any information present in the encoding of the initial value is lost. The legend shows the mean value for each delay conditioned distribution, \(\mu\), and the standard deviation, \(\sigma\).}
    \label{fig:final_values_drift_dataset}
    \Description[Final value resistance histogram.]{A histogram generated from the dataset of the final values attained after different delays for different initial starting resistances and delays.}
\end{figure}

A sampled subset of the series in the resistive drift dataset are shown in Figure~\ref{fig:drift_data}. We also show the output time series of the proposed model for the same initial resistance conditions for comparison.
The distribution of final resistance values across the dataset is shown in Figure~\ref{fig:final_values_drift_dataset}. We can see that over the course of 1000 seconds, almost all series converge to lie within a small region surrounding the equilibrium resistance of 500\(k\Omega\). This demonstrates that the information capacity of the device is a function of the delay, and decreases over time. Due to the drift towards the equilibrium point, all the information stored in the device will eventually be lost.
According to the modelling parameters chosen,
the energy imbalance between the rates of switching in the positive and negative directions, modulated by \(V_{\text{off}}\) (see \cite{el-geresyEventBasedSimulationStochastic2024})), results in an equilibrium resistance of 500\(k\Omega\), and thus, favours drift in the direction of the high resistance for resistances in the range chosen for the  dataset. This stochastic deviation of the resistance from its initial point is the distribution that we wish to model using the \ac{cGAN}.

\section{Conditional GAN (cGAN) Training}
\label{sec:method}
In this section, we introduce our approach to modelling the delay and initial resistance conditional distribution of resistance drift in memristors, through the use of time series generative modelling.
Letting \(r(t)\) denote the resistance as a random process, with \(t\) representing the time, our goal is as follows: we wish to find a computationally efficient and differentiable model that given an initial resistance value \(r(0)\) and a given delay \(d\), parameterises the conditional distribution \(p(r(d)|r(0), d)\).

\subsection{Delay Discriminator}
\label{sec:delay_discriminator}

As mentioned in Section~\ref{sec:time_series_generative_models}, several types of generative models have been proposed for modelling time series data. However, one major disadvantage to these approaches is that their output for a given timescale is fixed; this results in the need for recurrent evaluation of the models during inference. Thus the distribution \(p\left(r(d)|r(0),d\right)\) is decomposed into the factorised distribution based on the sampling step \(\Delta T\) of the time series generative model:

\begin{align}
    p(r(d)|r(0), d) = \prod_{i=0}^{(d/\Delta T) - 1} p\left(r\left((i+1)\cdot\Delta T\right)|r(i\cdot\Delta T)\right)
\end{align}

Since the purpose of our model is to be used as a proxy for a physical imperfection in an end-to-end differentiable learning setup, it is desirable to have a model that is as computationally simple and shallow (in terms of layers) as possible. The decomposition of the sampling into many steps can be both computationally demanding and can subsequently result in vanishing gradient issues in end-t-end settings. Additionally, if not appropriately addressed during training, issues can arise from a disconnect between closed-loop training and open-loop evaluation.

To address these problems, we propose to modify our generative modelling approach with the introduction of an additional condition: the delay, such that the final output, \(r(d)\), can be obtained directly by sampling the model output, that is - computing p(r(d)|r(0), d) directly. This will proffer several key advantages in our setting, and more generally:

\begin{enumerate}
    \item \textbf{Computational efficiency}: by conditioning directly on the delay, we are able to avoid the need for computationally expensive recurrent evaluation in order to generate outputs for larger delays. This reduction in computational complexity can also improve the flow of gradients through the model (fewer layers) when it is used in an end-to-end training setting.
    \item \textbf{Larger delays}: The model is able to learn to generate outputs conditioned on larger delays, by enforcing consistency with recurrent model output at smaller delays.
    \item \textbf{Addressing the disconnect between closed-loop training and open-loop evaluation}: By heavily reducing the need to recurrently run the model during inference, we minimise the possibility for open-loop evaluation errors to be introduced.
\end{enumerate}

Conditioning on the delay directly introduces additional complexity to our modelling problem, and makes it more difficult to train the model. To address this, we introduce a novel approach for enforcing consistency between time series generated at different resolutions, called delay discrimination. The adversarial approach relies on enforcing consistency in the generation of the outputs of the model, for the same total delay, evaluated at different timescales through an adversarial approach, where the model's own outputs at different scales are compared to one another (see Figure~\ref{fig:delay_comparison}).

We introduce a novel approach called \textit{delay discrimination}, which makes use of a network we call the \textit{delay discriminator}. This is an auxiliary discriminator network trained alongside the main generative model, forming an additional \ac{GAN}, with our generative model acting as the generator. We use this to improve the ability of the generative model to condition on the continuous delay conditions present in (that can be generated from) the training data, and also to improve the performance for larger delay conditions, outside of those available in the original training dataset.

We base our method on the inductive bias that a network conditioned on the delay should generate the same output distribution, regardless of the delay value, if evaluated recurrently, so long as the total delay is the same. For example, the network should generate the same conditional output distribution for a given initial resistance if it is conditioned directly on a total delay of 10, say, or if it used in a recurrent fashion for 10 iterations with a delay of 1. The delay discrimination algorithm is shown in Algorithm~\ref{alg:delay_discriminator} for a batch size of \(b\), where \(D_{dd ;\theta_{dd}}: \mathbb{R}^{2} \rightarrow [0, 1]\) represents the delay discriminator mapping, parameterised by \(\theta_{dd}\).

The principle is to train an auxiliary \ac{GAN} discriminator network to differentiate between the final values of two model series trained for different timescales. We first randomly generate a discriminator delay in the range \(d_{dd} \in [d_{ddmin}, d_{ddmax}]\), where \(d_{ddmin} > 0\) and \(d_{ddmax} \geq d_{max}\), as well as a random integer factor in the range \(q \in \{2, \ldots, q_{max}\}\). We then condition the generative model on the delay \(d_{dd}\), and given a randomly selected resistance \(r_{dd}\) as the initial condition, generate a single sample. Since this sample is generated with no reference to previously generated samples, it does not suffer from closed loop errors. Following this, we generate \(q\) samples using the generator, with a delay condition of \(d_{dd}/q\). We discard all but the conditioning resistance and the last sample for this second generated series. The delay discriminator, \(D_{dd ; \theta_{dd}}\), is trained to distinguish between final sample of the recurrent, open-loop evaluation for the smaller delay, and the sample generated using a single (non-recurrent) generation step for the larger delay. The result of this training process is to cause the generative model to change its generation of the series of delay \(d\) to be more akin to the series with delay \(d/q\), and vice versa. The procedure for training the delay discriminator is shown in Algorithm~\ref{alg:delay_discriminator}.
In practice, we use a value larger than 1.
Delay discrimination benefits the training in several ways:

\begin{enumerate}
    \item Information from the closed-loop evaluation for larger delays passes through to condition the open-loop evaluation for smaller delays, allowing inconsistencies to be highlighted and connecting the two forms of evaluation.
    \item Inclusion of the delay consistency inductive bias reduces training complexity and improves convergence.
    \item It allows data to be generated with delays larger than present in the dataset, as well as fractional delays, allowing for a much wider range of delays to be used for conditioning the generator in the training process. Following training, the generator can model the conditional distribution after much larger delays, without many recurrent evaluation steps.
\end{enumerate}

\begin{algorithm}[t]
	\caption{Algorithm for updating the network parameters of the generator and delay discriminator for a single batch of size \(b\). \(\alpha\) is the learning rate.}
	\label{alg:delay_discriminator}
	\begin{algorithmic}[1]
        \State $q \sim U\{2, q_{\text{max}}\}$
        \State $\mathcal{P}_{dd, 1} \gets \{\}$
        \State $\mathcal{P}_{dd, 2} \gets \{\}$
        \For{$i \in b$}
            \State $d_{dd} \in U\{d_{\text{min}, dd}, d_{\text{max}, dd}\}$
            \State $r_{\text{init}} \gets U\{R_{\text{min}}, R_{\text{max}}\}$
            \State $r_{\text{final}, 1} = G_{\theta_g}(r_{\text{init}}, d_{dd})$
            \State $r_{\text{final}, 2} \gets r_{\text{init}}$
            \For{$j \in q$}
                \State $ r_{\text{final}, 2} \gets G_{\theta_g}(r_{\text{init}}, d_{dd}/q) $
            \EndFor
            \State  $\mathcal{P}_{dd, 1} \gets \mathcal{P}_{dd, 1} \cup D_{dd ; \theta_{dd}}(r_{\text{init}}, r_{\text{final}, 1})$
            \State $\mathcal{P}_{dd, 2} \gets \mathcal{P}_{dd, 2} \cup D_{dd ; \theta_{dd}}(r_{\text{init}}, r_{\text{final}, 2})$
        \EndFor
        \State $\theta_{dd} \gets \theta_{dd} + \alpha\nabla_{\theta_{dd}}(BCE(\mathcal{P}_{dd, 1}, \{0, \ldots 0\}) + \alpha\nabla_{\theta_{dd}}(\mathcal{P}_{dd, 2}, \{1, \ldots 1\})$
        \State $\theta_{g} \gets \theta_{g} + \alpha\nabla_{\theta_{dd}}(BCE(\mathcal{P}_{dd, 1}), \{1, \ldots 1\}))$
    \end{algorithmic}
\end{algorithm}

We will highlight the effectiveness of the delay discriminator method in Section~\ref{sec:evaluation_visual}, where we look at the mean and variance of the values after an equivalent total delay of 1000 for series generated using delay conditions in the set \([5, 10, 100, 250, 1000]\). We will demonstrate that using the delay discriminator vastly improves consistency of generation, especially for larger delays.

\subsection{Data Transformation and Normalisation}

For any data-driven optimisation problem, it is important to appropriately transform the input data, in order to incorporate known inductive biases that facilitate learning. It is also well-known that data normalisation aids the learning process by ensuring that there is no bias in the contribution of features to the gradients of the network during optimisation.

The nature of our input features - the initial resistances and the delays - presents a challenge to our learning algorithm, due to their domain spanning multiple orders of magnitude. We must choose an appropriate data transformation that reduces the large range, while also allowing for precision in generation at a variety of scales.

In the case of the normalisation of the resistance values, we apply a logarithmic transform followed by subtraction of the mean and scaling by the standard deviation (using statistics empirically averaged over the drift dataset) in order to normalise the magnitudes. We determine the parameter \(\mu_R\) as the sample mean of the logarithmically transformed resistances, taken over the dataset, and similarly find \(\sigma_R\) as the square root of the sample variance of the resistances present in all resistance series, \(S_i\) - indexed by \(i\), in the dataset, \(\mathcal{D}\):
\begin{align}
    \notag
    \mu_R &= \frac{1}{|\mathcal{D}|}\sum_{S_i \in \mathcal{D}}\frac{1}{|S_i|}\sum_{r_j^i \in S_i} \ln{(r_j^i)} \; ,\\
    \sigma_R &= \sqrt{ \frac{1}{|\mathcal{D}|}\sum_{S_i \in \mathcal{D}}\frac{1}{|S_i|}\sum_{r_j^i \in S_i} (\ln{(r_j^i)} - \mu_R) } \; ,
    \label{eq:dataset_std}
\end{align}
where \(r_j^i \in S_i\) are the ordered (resistance) elements of the series \(S_i\), indexed by \(j\). Thus, the resistance normalisation transformation, \(N_{\text{res}}(r)\), is given as:
\begin{align}
    N_{\text{res}}(r) = \frac{\ln{(r)}-\mu_R}{\sigma_R}
    \label{eq:data_transform}
\end{align}

We found that it was more efficient to train the generator network to predict the differences in the normalised resistance values, rather than the absolute values. As such, we apply a residual connection \cite{heDeepResidualLearning2016} between the input and the output. Prediction of the difference incorporates the inductive bias that the output should be equal to the input by default for a value of 0. This inductive bias is a common feature of popular neural network architectures, such as U-Net \cite{ronnebergerUNetConvolutionalNetworks2015} or multi-head attention modules in transformers \cite{vaswaniAttentionAllYou2017}, which use residual connections.

To normalise the differences of the normalised dataset, we apply a conditional mapping - \(N_{\text{diff}}(\cdot, \cdot)\) - that is dependent on both the normalised input resistance, \(\bar{r}_{\text{init}} = N_{\text{res}}(r_{\text{init}})\), and the normalised output resistance, \(\bar{r}_{\text{final}} = N_{\text{res}}(r_{\text{final}})\), resulting in the normalised difference, denoted by \(\bar{r}_{\text{diff}}\). Note that the difference normalisation transform is applied atop data normalisation. Let \(\bar{\mathcal{D}}\) denote the normalised dataset, created by applying \(N_{\text{res}}(\cdot)\) pointwise to all \(r_j^i \in S_i \; \forall  S_i \in \mathcal{D}\). We denote the series of the normalised dataset as \(\bar{S}_i \in \bar{\mathcal{D}}\) and the resistance values in each normalised series as \(\bar{r}_j \in \bar{S}_i\). The forward (Equation~(\ref{eq:forward_differences_transform})) and inverse (Equation~(\ref{eq:inverse_differences_transform})) difference normalisation transforms are defined as follows, with the standard deviation used for normalisation calculated from the dataset statistics for a delay of 1:

\begin{align}
    \notag &\mu_{\bar{\mathcal{D}}} = \frac{1}{|\bar{\mathcal{D}}|}\sum_{S_i \in \bar{\mathcal{D}}}\frac{1}{|S_i|-1}\sum_{\{\bar{r}_j^i \in \bar{S}_i | j < |\bar{S}_i|\}} (\bar{r}_{j+1}^i - \bar{r}_{j}^i) \\
    \notag &\sigma_{\bar{\mathcal{D}}} = \sqrt{ \frac{1}{|\bar{\mathcal{D}}|}\sum_{\bar{S}_i \in \bar{\mathcal{D}}}\frac{1}{|\bar{S}_i|-1}\sum_{\bar{r}_j^i \in \bar{S}_i} (\bar{r}_j^i - \mu_{\bar{\mathcal{D}}}) } \\
    &N_{\text{diff}}^{-1}(\bar{r}_{\text{diff}}, \bar{r}_{\text{init}}) = \bar{r}_{\text{init}} + \bar{r}_{\text{diff}} \cdot \sigma_{\bar{\mathcal{D}}}
    \label{eq:inverse_differences_transform} \\
    &N_{\text{diff}}(\bar{r}_{\text{final}}, \bar{r}_{\text{init}}) = \frac{(\bar{r}_{\text{final}} - \bar{r}_{\text{init}})}{\sigma_{\bar{\mathcal{D}}}}
    \label{eq:forward_differences_transform}
\end{align}

\begin{figure}
    \centering
    \hspace*{\fill}
    \subfloat[Histogram of the normalised resistances, obtained by applying the transform \(N_{\text{Res}}(\cdot)\) pointwise to the resistance values in the series in the dataset \(D\).]{
        \includegraphics[width=0.45\textwidth]{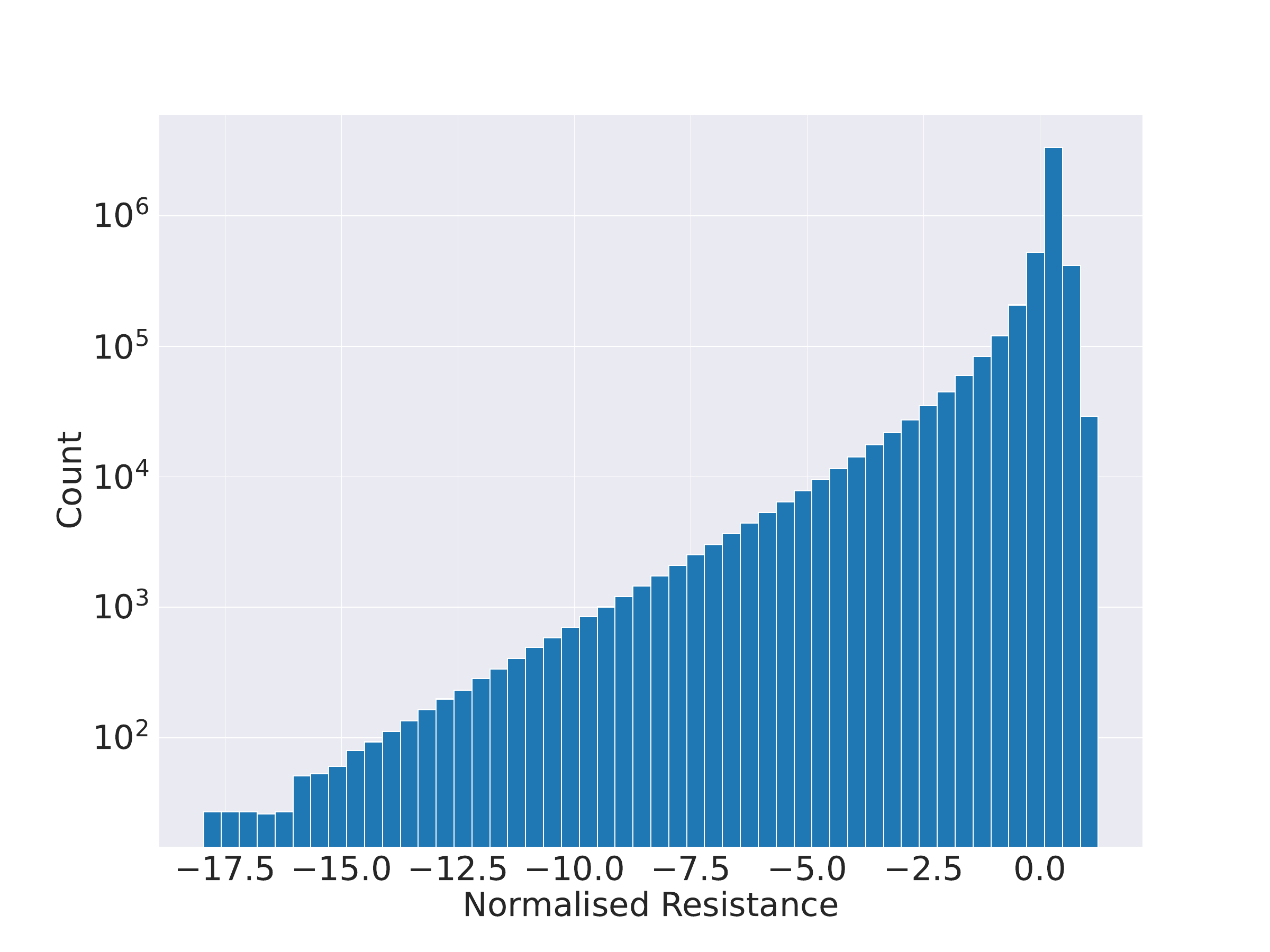}
        \label{fig:chapsorage_normalisation_transforms_series}}
        \hfill
    \subfloat[Normalised histograms for the differences, for a range of delay conditions, derived empirically from the dataset \(D\), obtained by applying the difference normalisation transform \(N_{\text{diff}}(\cdot, \cdot)\) to the pairs of values separated by delay \(\Delta T\).]{
        \includegraphics[width=0.45\textwidth]{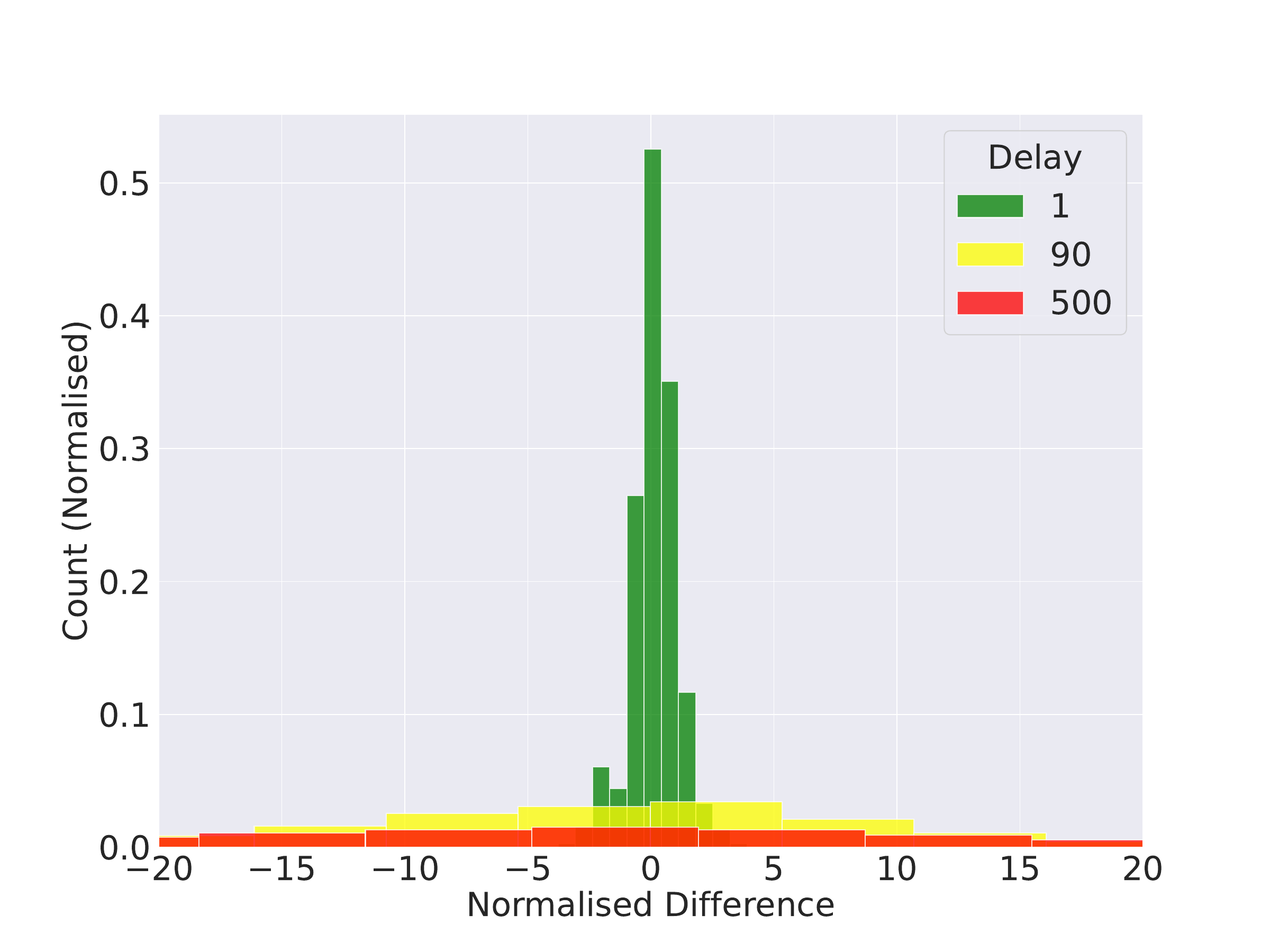}
        \label{fig:chapsorage_normalisation_transforms_differences}}
    \hspace*{\fill}
    \caption{Histograms of the values in the dataset, \(D\) following the initial pointwise application of the resistance normalisation transform (Figure~\ref{fig:chapsorage_normalisation_transforms_series}) and subsequent application of the difference normalisation transform for pairs of values from the dataset separated by a given delay (Figure~\ref{fig:chapsorage_normalisation_transforms_differences}).}
    \label{fig:normalisation_transforms}
    \Description[Histograms of the normalised values from the dataset.]{Two histograms generated from the data in the dataset. One showing the normalised resistance values, and the other showing the normalised difference values.}
\end{figure}

Figure~\ref{fig:normalisation_transforms} shows the distribution of values following the application of the resistance normalisation and difference normalisation transforms to the resistances in the dataset. In both cases, the normalisation transforms map the data to a more reasonable input range, while also retaining precision in their descriptive capabilities at different orders of magnitude. Note that, in the case of the difference normalisation transform, larger delays result in a larger range of values observed in the transformed histogram. It is also possible to ensure that the histograms are limited in their range by incorporating delay scaling in the difference normalisation transform; however, we found that retaining the different ranges of the transformed resistances for different delay conditions actually improved the training performance.

Similarly, despite the large range of delays used for training, denoted by \( [d_{\text{min}}, d_{\text{max}}]\) with \(d_{\text{min}}=1\) and \(d_{\text{max}}=500\), it was found that scaling of the delays using a logarithmic transform did not benefit training. This is likely due to the magnitude of the normalised differences (the feature that the generator is trained to predict) being somewhat proportional to the delay condition. Thus, we apply no transformation to the delays and pass the unscaled delays directly to the network.

\subsection{cGAN Generative Model}

\acp{GAN} are generative models that have been shown to be capable of effectively modelling complex distributions through the use of adversarial training \cite{goodfellowGenerativeAdversarialNetworks2014}. We will use a conditional form of \acp{GAN}, called \acp{cGAN} \cite{mirzaConditionalGenerativeAdversarial2014}, where the generator parameterises the delay and initial resistance conditioned resistive drift distribution. Time Series \ac{GAN} is one \ac{GAN} that has shown success in generating realistic time series data \cite{yoonTimeseriesGenerativeAdversarial2019}; however, the model still requires recurrent evaluation for the generation of time series values at arbitrary points in the future.

We use a canonical \ac{BCE} loss for the discriminator for our \ac{GAN} training. We denote the generator network as \(G_{\theta_g}\), parameterised by \(\theta_g\); and we denote the main discriminator network as \(D_{d;\theta_d}\), parameterised by \(\theta_d\) \cite{goodfellowGenerativeAdversarialNetworks2014}. Let us denote the distribution induced by sampling from \(G_{\theta_g}\) as \(p_{f}(f)\), and the ground truth data distribution as \(p_{x}(x)\). Our discriminator objective is the maximisation of the likelihood of classifying samples as \(1\), or ``real'', when evaluated on real data samples, and the simultaneous maximisation of the likelihood of classifying samples as \(0\), or ``fake'', when evaluated on the distribution induced by the generator. This is achieved by the discriminator's minimisation objective:

\begin{align}
    L_{\theta_d} = &- \mathbb{E}_{x \sim p_{x}(x)}{\left[ \log{(D_{d;\theta_d}(x))} \right]} \nonumber \\
    &- \mathbb{E}_{f \sim p_{F}(f)}{\left[ \log{(1 - D_{d;\theta_d}(f))} \right]}
    \label{eq:discriminator_objective}
\end{align}

Our generator objective is to maximise the likelihood of the discriminator classifying samples generated from the induced distribution as \(1\), or ``real''. This is achieved by the generator's minimisation objective being:

\begin{align}
    L_{\theta_d} = - \mathbb{E}_{x \sim p_{x}(x)}{\left[ \log{(D_{d;\theta_d}(x))} \right]}
    \label{eq:generator_objective}
\end{align}

Wasserstein \ac{GAN} proposes an alternative optimisation objective, requiring the use of the associated techniques of discriminator weight or gradient clipping to enforce the Lipschitz constraints on the discriminator \cite{arjovskyWassersteinGAN2017}. However, we found that for our problem setting, the original \ac{GAN} loss function as described in Equations~(\ref{eq:discriminator_objective})~and~(\ref{eq:generator_objective}) above, yielded better convergence.

\subsection{Training Details}

\subsubsection{Architecture}

\begin{figure}
    \centering
    \includegraphics[width=\linewidth]{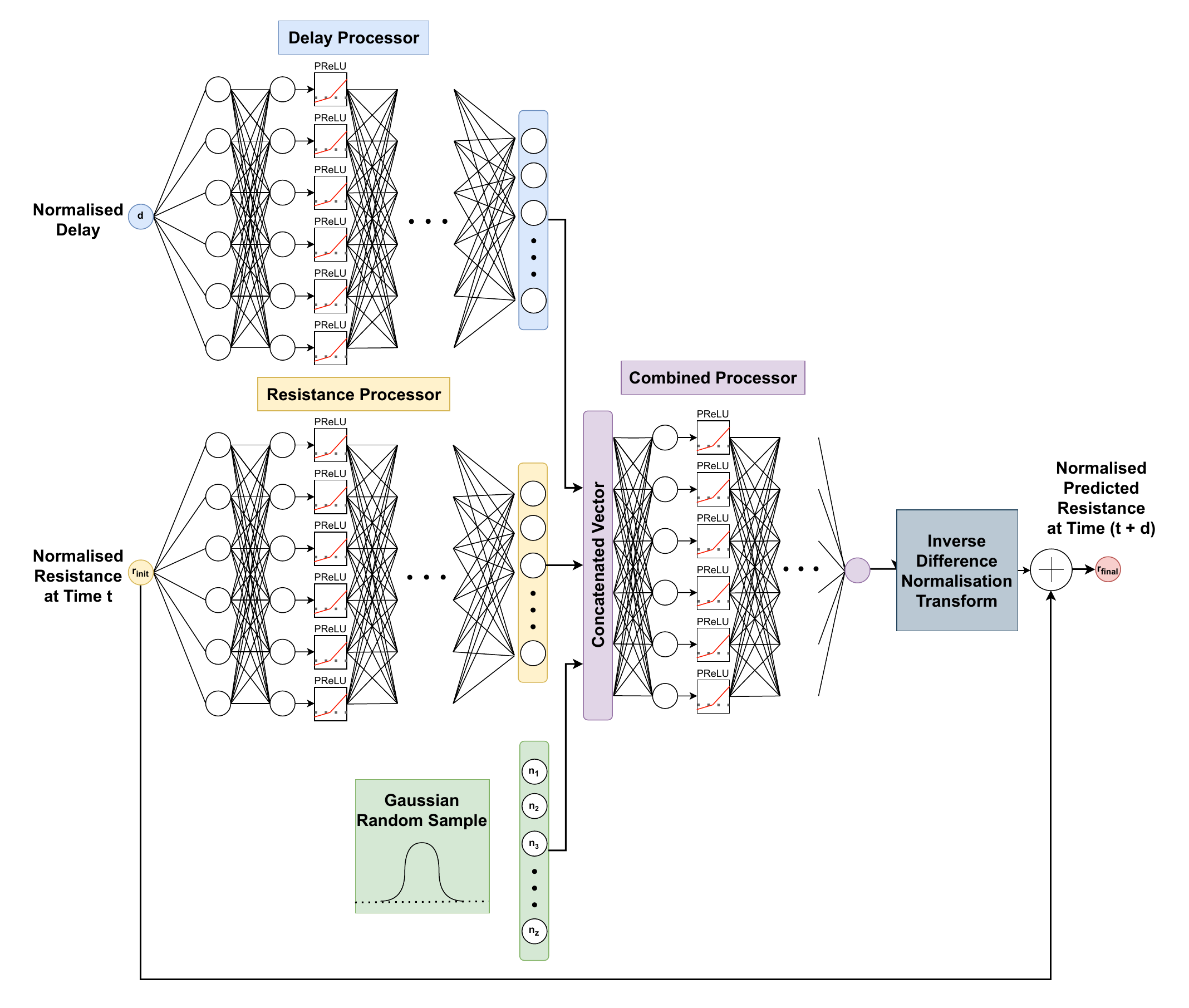}
    \caption{Structure of the \ac{cGAN} used for training. The delay processor and resistance processor map the delay \(d\) and the normalised resistance at time \(t\), denoted by \(\bar{r}_{\text{init}}\), to independent embeddings. The combined processor network then processes these embeddings along with the diagonal Gaussian noise vector of dimension \(z\), which is the latent prior for the generative model. The output of the combined processor is a normalised difference, to which the inverse difference normalisation transform is applied, followed by addition to \(\bar{r}_{\text{init}}\) to produce the predicted resistance after delay \(d\), \(\bar{r}_{\text{final}}\).}
    \label{fig:cgan_structure}
    \Description[Structure of the conditional generative adversarial network.]{A figure showing the structure of the conditional generative adversarial network, with a separate delay and initial resistance processing module, as well as a noise vector input. The output is the predicted resistance after delay d.}
\end{figure}

\begin{figure}
    \centering
    \includegraphics[width=\linewidth]{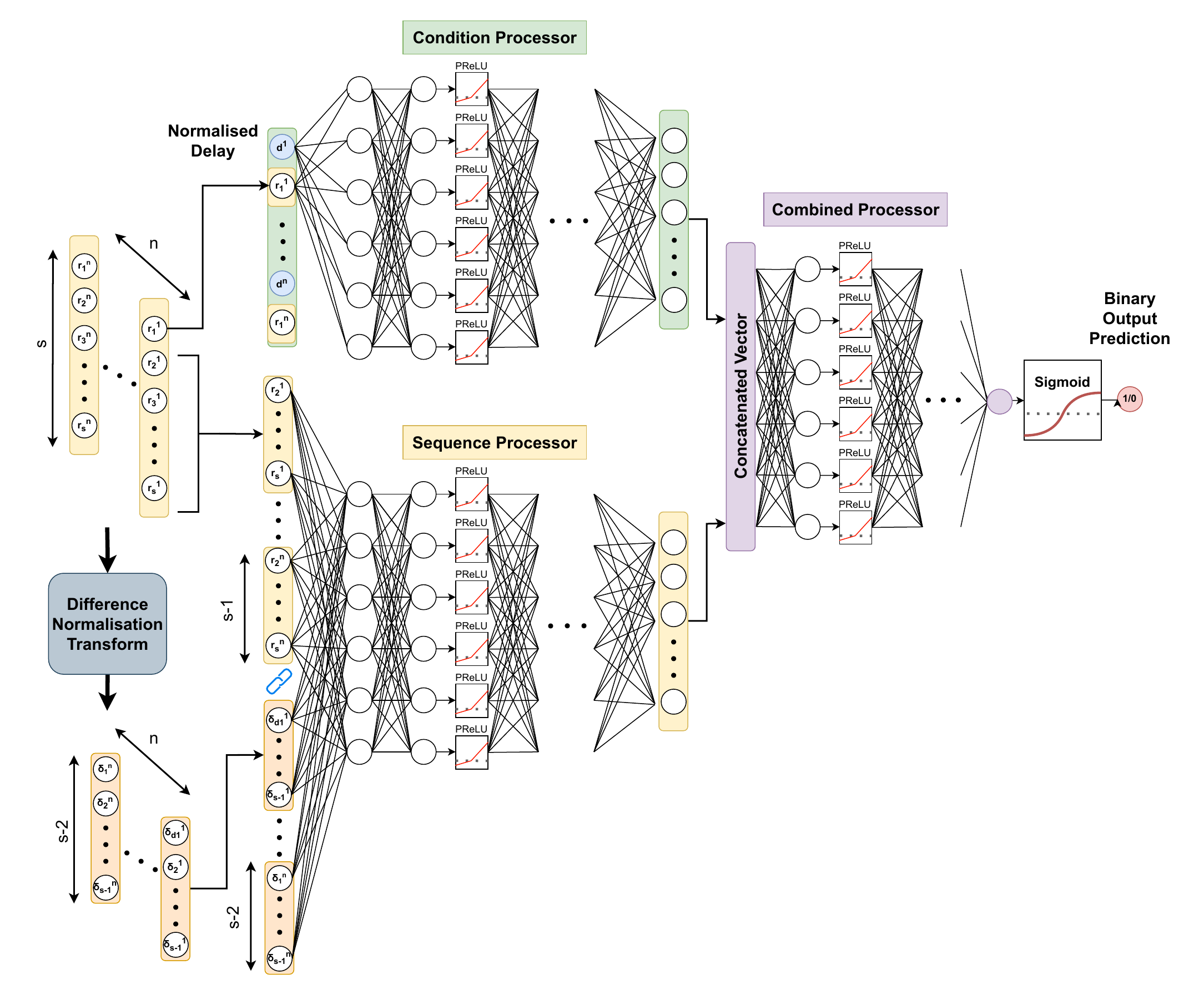}
    \caption{The discriminator architecture and the multiple sample discrimination approach. The condition processor processes the first element of the sequence, i.e., the initial condition, in addition to the delay. The remaining elements of the sequence vector: \(r_2 \ldots r_s\), are passed to the sequence processor, along with the output of the difference normalisation transform applied to each sequence. The output embeddings of both processors are concatenated and passed to the combined processor. When using the technique of multiple sample discrimination, we pass \(n\) sequences (and \(n\) delays) to the network simultaneously, with the condition processor and sequence processor input dimensions being scaled up by \(n\) and the effective batch size for the discriminator scaled down by \(n\). The same architecture is used for the main and auxiliary delay discriminators described in Section~\ref{sec:delay_discriminator}.}
    \label{fig:discriminator}
    \Description[Structure of the discriminator architecture.]{The discriminator architecture is shown along with the process behind the multiple sample discrimination approach. The condition processor module takes in multiple generated or real resistance series as well as appropriate versions with the normalisation transforms applied. The output is a binary real or fake prediction.}
\end{figure}

\begin{figure}
    \centering
    \includegraphics[width=0.9\linewidth]{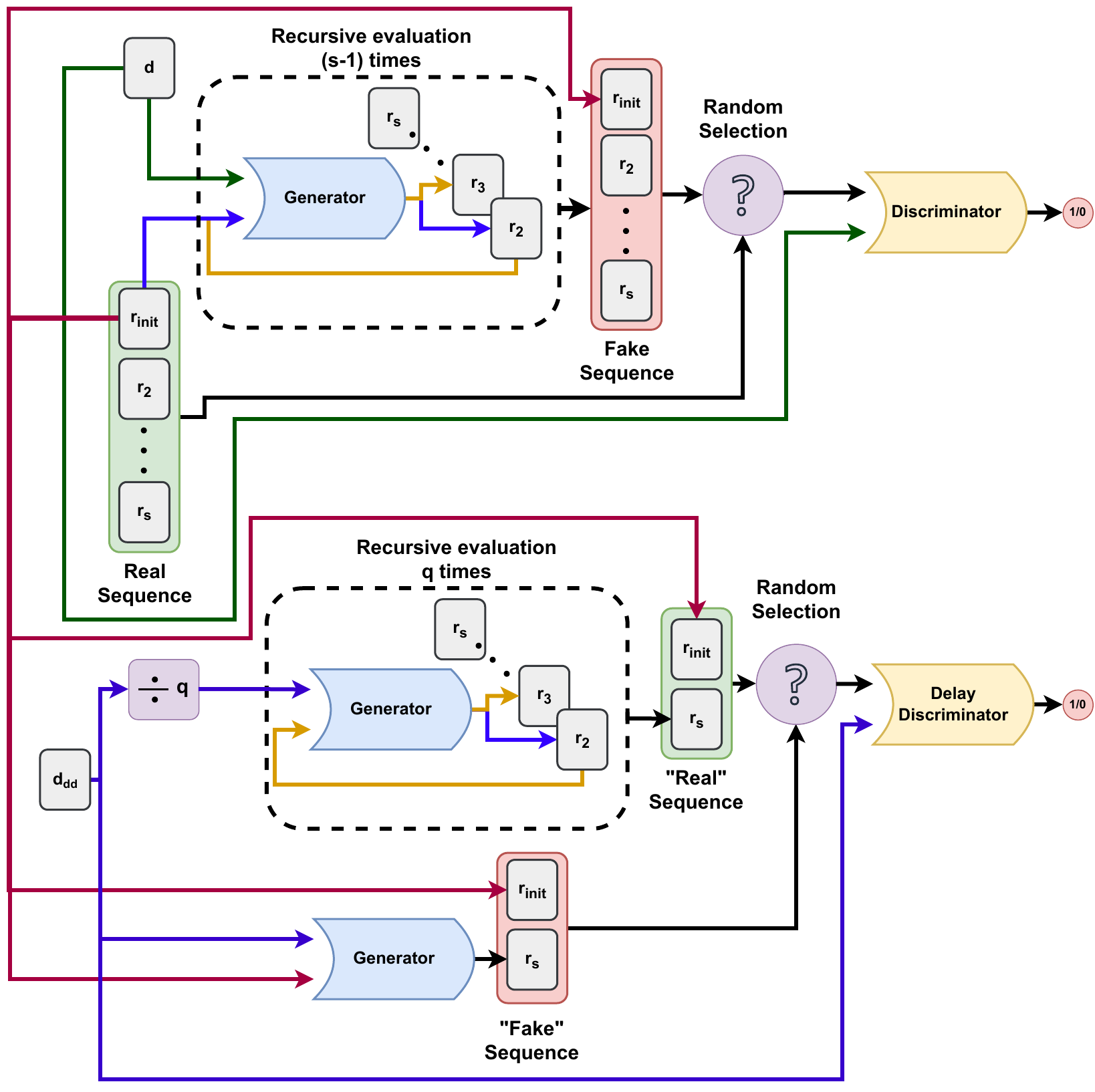}
    \caption{The training structure for the generator and the two discriminators. This visualises the delay discrimination procedure. For the purposes of clarity in visualisation, we do not show multiple sample discrimination in this figure. The top of the diagram shows the generation of a fake sequence for the canonical adversarial training, with the discriminator trained to distinguish between real and fake resistance sequences. The bottom of the diagram shows the training of the \ac{cGAN} to enforce consistency in its output for the same total delay, also addressing the problems with closed loop training described in Section~\ref{sec:delay_discriminator}.}
    \label{fig:training_structure}
    \Description[Overall structure of the GAN training, including the proposed delay discrimination approach.]{The overall training structure is shown, with the generator and discriminator models being trained adversarially. The delay discriminator network aids in the training process by enforcing consistency between generated values at different timescales.}
\end{figure}

We use a modular, fully connected architecture, with ReLU non-linear activation functions for both the generator and the discriminator. We observed that, in practice, using batch normalisation in the generator network harmed the performance.
A diagram showing the network architecture for the generator network is shown in Figure~\ref{fig:cgan_structure}. The discriminator network architecture is shown in Figure~\ref{fig:discriminator}. A network diagram, showing the overall training procedure, including both the discriminator and the auxiliary delay discriminator, is shown in Figure~\ref{fig:training_structure}.

\subsubsection{Feeding Data to the Network}

We train in a mixed-mode setup, with the normalised initial resistance condition being passed as an initial value and the subsequent values fed recurrently back to the network to generate further samples of the series. Once a sequence of \(s-1\) additional values have been recurrently generated by the generator network, the entire generated sequence is fed simultaneously to the discriminator. This ensures that both open-loop and closed-loop errors (see Section~\ref{sec:time_series_generative_models}) are used simultaneously to update the network parameters, allowing for the network to be used recurrently if needed during evaluation.

We experimented with the choice of which transformed versions of the input to pass to each of the networks - the generator and the discriminator(s). We found that presenting a vector consisting of both the normalised resistances, \(\bar{r}_{\text{init}}\), and the normalised differences, \(\bar{r}_{\text{diff}}\), to the discriminator improved the performance.
The normalised resistances can be seen as providing information about the absolute value of the input (the order of magnitude), while the normalised differences provide more fine-tuned information about the relative values. The discriminator is fed a series of \(s\) normalised resistance values simultaneously, along with a series of \(s-1\) normalised difference values, that are obtained by computing the difference normalisation transform of the pairs of resistance values. The first element of the resistance series fed to the discriminator is always a real datapoint, and the other \(s-1\) values are either the remainder of the series generated from the training set, or \(s-1\) generated values. The output of the discriminator is a real number in the range \([0, 1]\) representing its belief that the given resistance and delay values were drawn from the dataset (1) or that they were generated (0). The dimensionality and the nature of the inputs to the discriminator is visualised in Figure~\ref{fig:discriminator}.

Although presenting the normalised differences to the discriminator resulted in improvements to the training, we did not pass the normalised resistances as inputs to the generator, in order to alleviate the need for more than one initial data point to condition the generator, allowing it to output a sample from a conditional distribution (and subsequent recurrently evaluated samples) given a single resistance conditioning value \(r_{\text{init}}\). In the generation of resistance series, we use no state variable, conditioning the generator only on the previous resistance value, implicitly incorporating the inductive bias that the resistance series are memoryless. The generator input is thus a pair of scalar variables representing a normalised resistance and a delay condition. Respectively, we apply an inverse difference normalisation transform at the output of the generator, along with a residual connection to add the resultant difference to the normalised input resistance to produce a normalised generated final resistance. The dimensionality and nature of the inputs to the generator can be seen in Figure~\ref{fig:cgan_structure}.

\subsubsection{Multiple Sample Discrimination}
\label{sec:multibatch_discrimination}

One notable challenge that plagues the training of \acp{GAN} is the problem of mode collapse \cite{arjovskyPrincipledMethodsTraining2016}. In our case, the problem presents itself as a tendency for the network to accurately learn the mean value of the resistance changes, but to learn a variance that is smaller than that of the true conditional distribution.

In order to improve the discriminator's effectiveness at identifying this phenomenon in an attempt to alleviate mode collapse, we pass \(n\) real or generated sequences jointly to the discriminator as input. Thus, our effective batch size is reduced from its initial value \(b\) to \(b/n\) for the discriminator. This technique has been previously demonstrated to improve training by addressing the problem of mode collapse \cite{linPacGANPowerTwo2020, salimansImprovedTechniquesTraining2016}.

The disadvantage of this technique, in addition to the reduction in the size of the batch, is that it increases the effective dimensionality of the discriminator input. We must therefore trade-off these disadvantages for the reduced level of mode-collapse observed with this increased computational complexity. In practice, we find that a value of \(n=2\) produces good results and that is the setting used for the results presented in Section~\ref{sec:results}.

This technique is applied to both the main discriminator network, and the auxiliary delay discriminator network, described in  Section~\ref{sec:delay_discriminator}. The use of multiple simultaneous samples for input to the discriminator is shown in a diagram depicting the discriminator architecture in Figure~\ref{fig:discriminator}.

\subsubsection{Parameters}

We optimise all networks using an Adam optimiser \cite{kingmaAdamMethodStochastic2017} with a learning rate of \(1 \times 10^{-4}\). We set the sequence length for the generation, and also thus the number of resistance values passed to the discriminator in a single sequence, as \(s=10\). We set the sequence length for input to the delay discriminator as \(s=2\) (i.e., a single generation) although, as described in Algorithm~\ref{alg:delay_discriminator}, the generator is run with either a sequence length of \(s=2\) or a sequence length of \(s=1+q\) in the case of the delay discriminator, for both the full and split delay settings.

We run each \ac{GAN} training experiment for 1000 epochs, with each epoch consisting of 500 steps, where all three networks are updated simultaneously. We set the value of \(q_{\text{max}} = 20\). The dimension of the latent prior for the generator, denoted by \(z\), is set to 20. We set the minimum and maximum delays for the main discriminator as \(d_{\text{min}, d} = 1\) and \(d_{\text{max}, d} = 90\), and those for the delay discriminator as \(d_{\text{min}, dd} = 1\) and \(d_{\text{max}, dd} = 500\).

\section{Results}
\label{sec:results}
We evaluate the distribution matching performance of the trained generator model, compared to the true distribution of the data in the resistive drift dataset, in several different ways. In order to highlight the consistency of the model output across delays, whose improvement is a key benefit of the delay discriminator approach, we evaluate across a range of equivalent delays using a variety of smaller timesteps. We can also visually compare the series output by the model during closed loop evaluation, which can highlight problems associated with the compounding of closed loop errors and can give some sense of the quality of the generative model. Finally, we compare the statistics of the \ac{cGAN} model output to those of the ground truth event-based model. We statistically estimate the mean and variance (the first and second order statistical moments) of the generated data, as well as showing empirical model outputs for a variety of combinations of delay and initial resistance conditions.

In order to measure the impact of the techniques used to improve the model training, during evaluation, we also perform an ablation experiment which does not make use of the delay discriminator.

\subsection{Delay Consistency Evaluation}

\begin{figure}[h]
\centering
\hspace*{\fill}
\subfloat[Mean comparison.]{
    \includegraphics[width=0.45\textwidth]{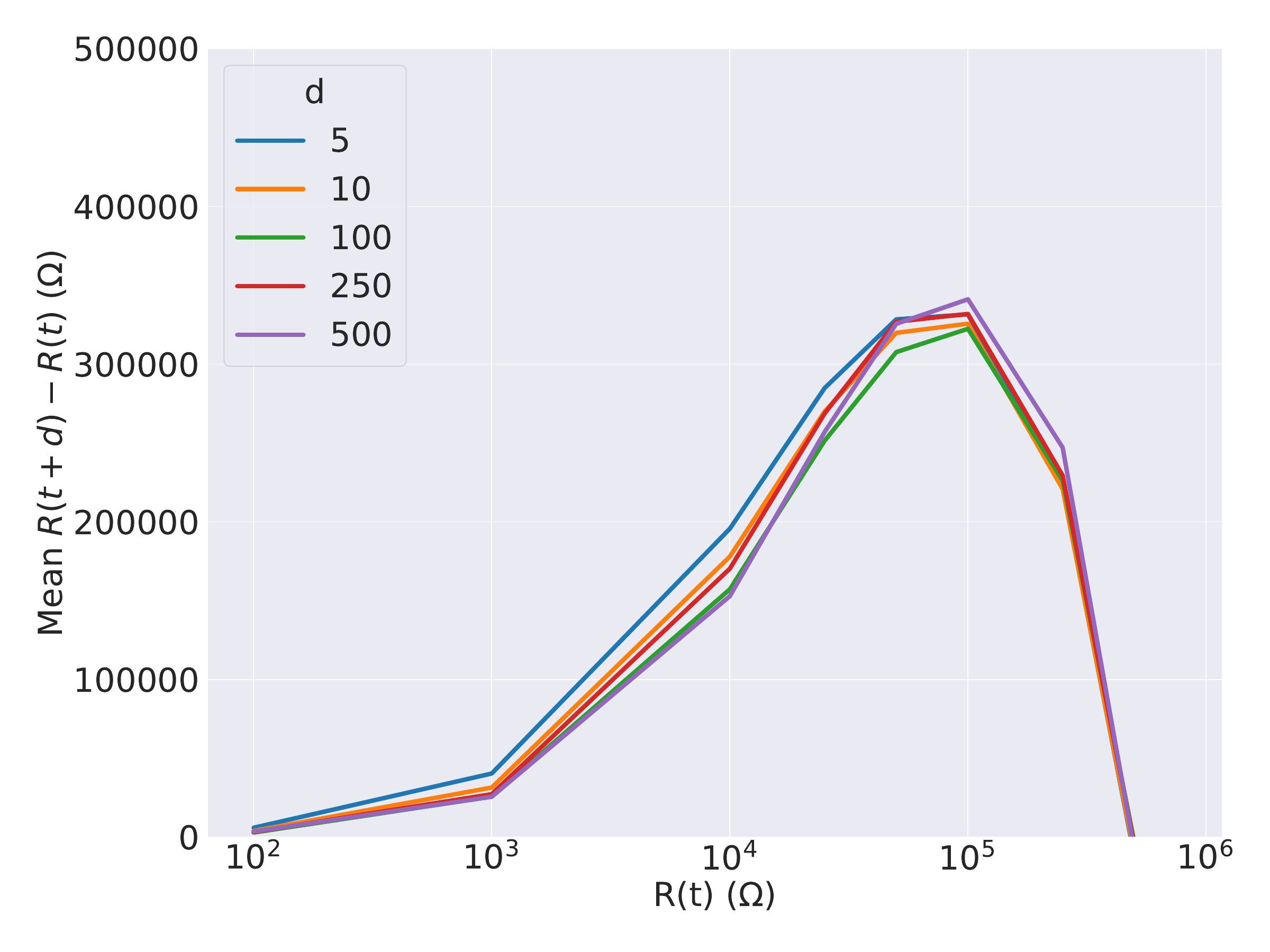}
    \label{fig:delay_comparison_means_real}
}
    \hfill
\subfloat[Standard deviation comparison.]{
    \includegraphics[width=0.45\textwidth]{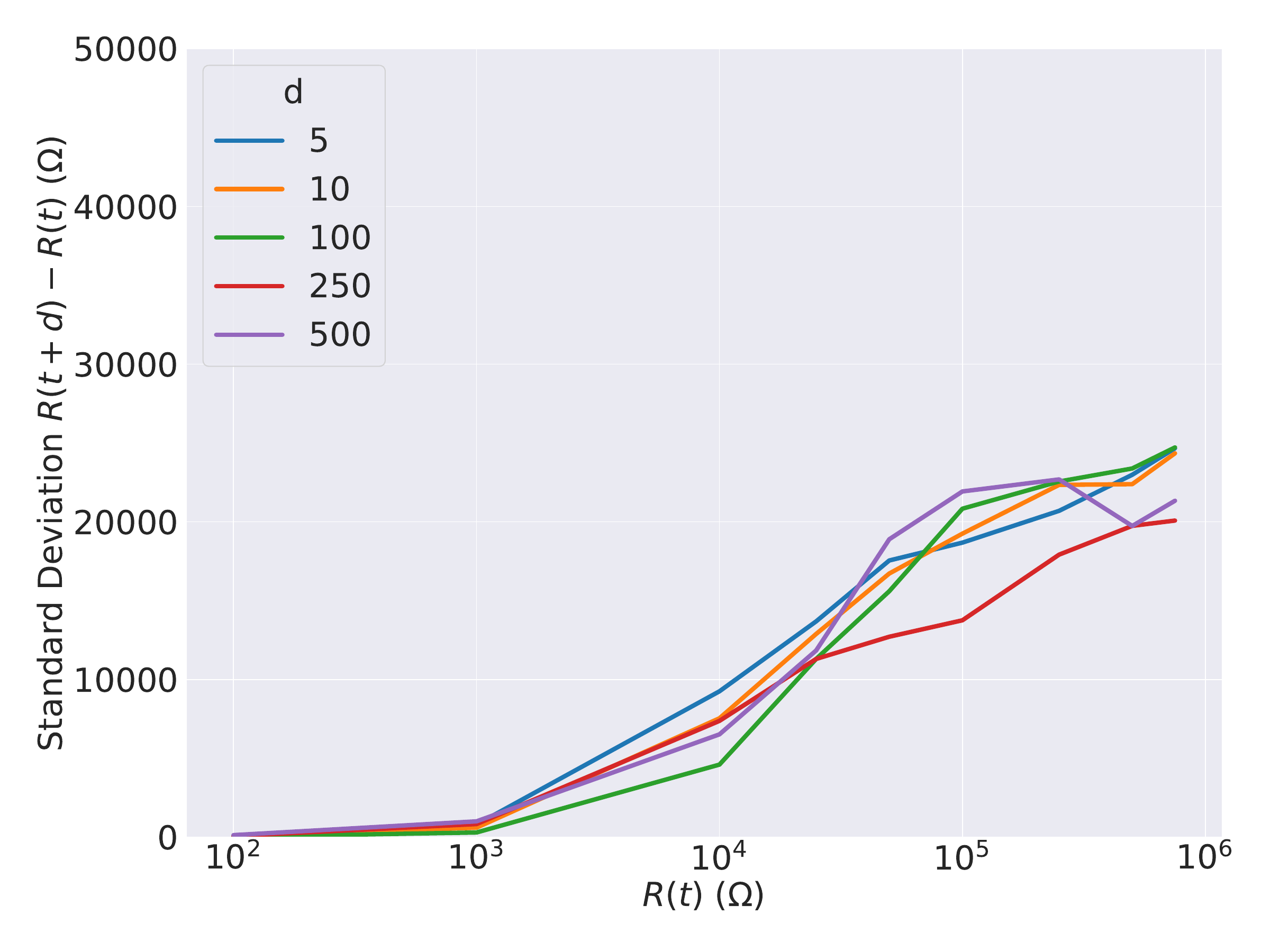}
    \label{fig:delay_comparison_stds_real}}
    \hspace*{\fill}
    \\
    \hspace*{\fill}
\subfloat[Mean comparison for delay discriminator ablation experiment.]{
    \includegraphics[width=0.45\textwidth]{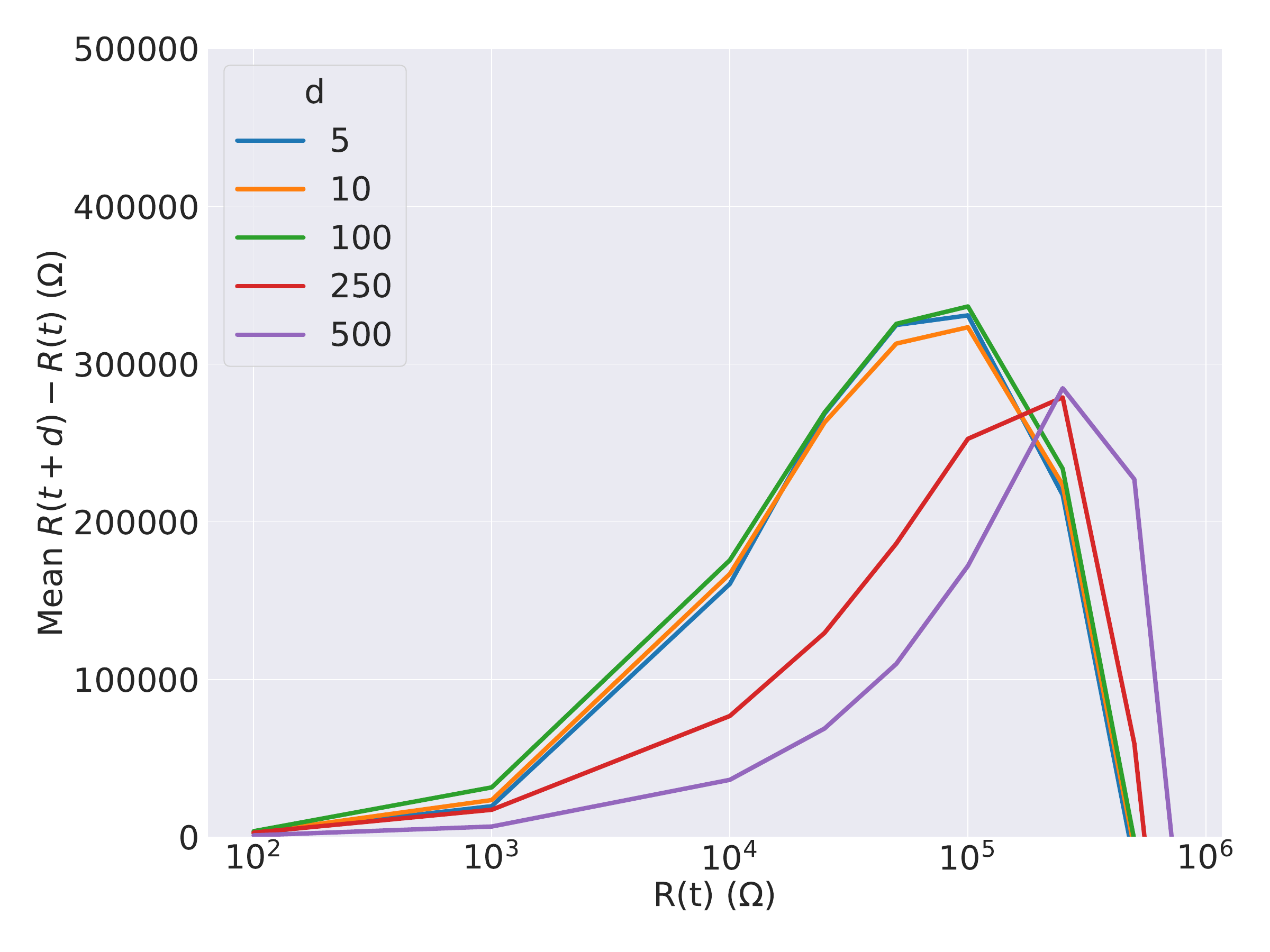}
    \label{fig:delay_comparison_means_ablation}}
    \hfill
\subfloat[Standard deviation comparison for delay discriminator ablation experiment.]{
    \includegraphics[width=0.45\textwidth]{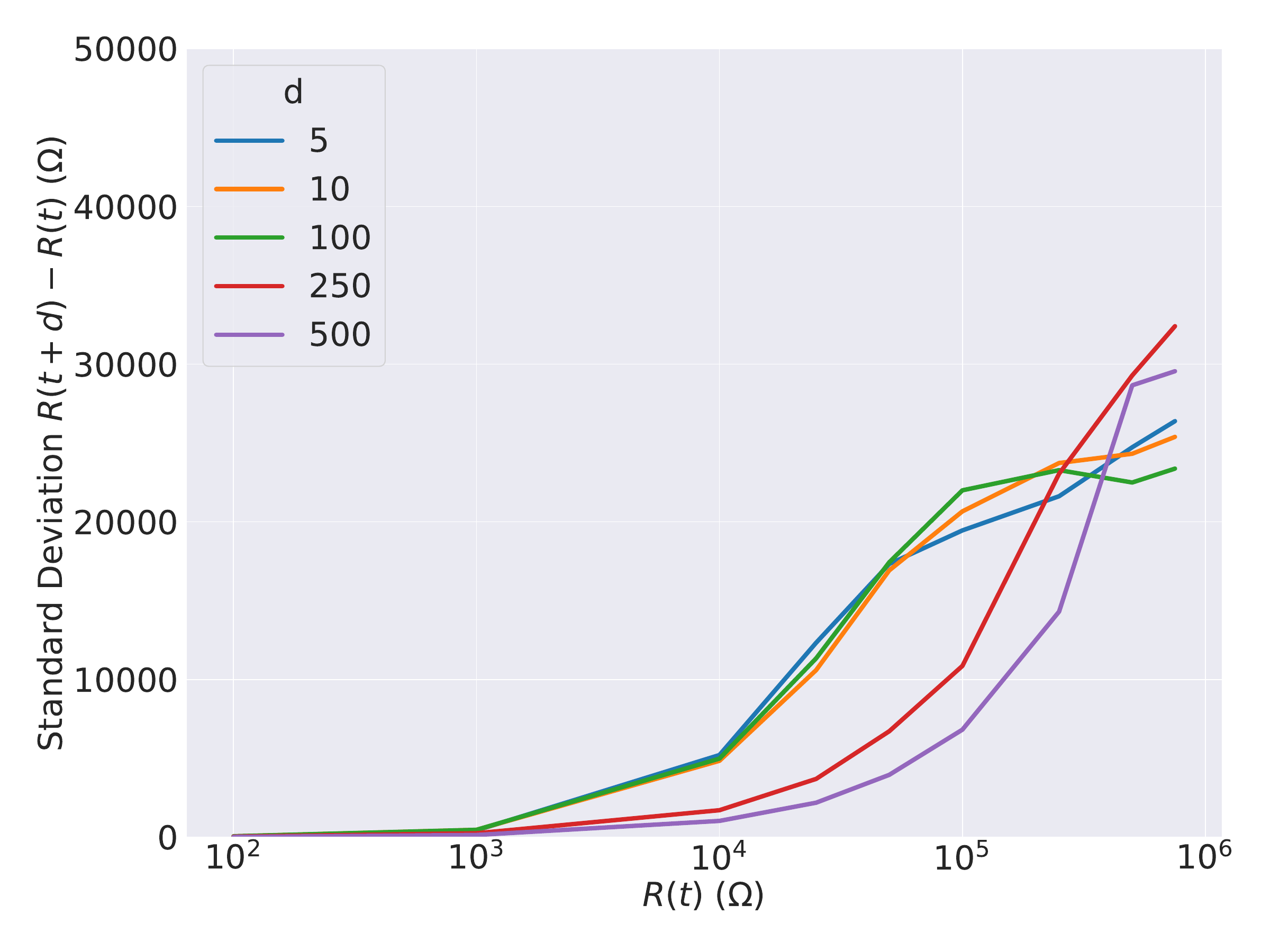}
    \label{fig:delay_comparison_stds_ablation}}
    \hspace*{\fill}
\caption{Experiments examining the consistency of the model for generation of outputs of the same cumulative delay, \(\Delta T\), at different levels of delay refinement \(d\). We measure the mean change in the resistance for a variety of starting resistance values. In the case of a delay of \(d\), we use \(500/d\) steps run in order to achieve a total equivalent delay of \(\Delta T = 500\).}
\label{fig:delay_comparison}
\Description[A comparison of the model output for different delay conditions.]{The statistics of the generative model's output are evaluated for different delay conditions, for both the proposed modelling approach, and an ablation experiment run without the delay discriminator, showing the advantages of the proposed delay discrimination method.}
\end{figure}

Here, we compare the mean difference from the initial resistance at time \(t=0\), for the model run for an equivalent delay of 500, generated by conditioning with a delay condition of (5, 10, 100, 250, and 500), for (1, 2, 5, 50, and 100) recurrent steps, respectively, in order to achieve the same total delay for the model's final resistance output. We see that the model's output is relatively consistent across delays. We evaluate this behaviour for experiments run with, and without, the delay discriminator. We find that the delay discriminator improves the consistency of the model outputs for the same total delay for different given delay conditions.
This is shown in Figure~\ref{fig:delay_comparison}.

\subsection{Qualitative Evaluation}
\label{sec:evaluation_visual}

\begin{figure}[h]
\centering
\hspace*{\fill}
\subfloat[]{
    \includegraphics[width=0.3\textwidth]{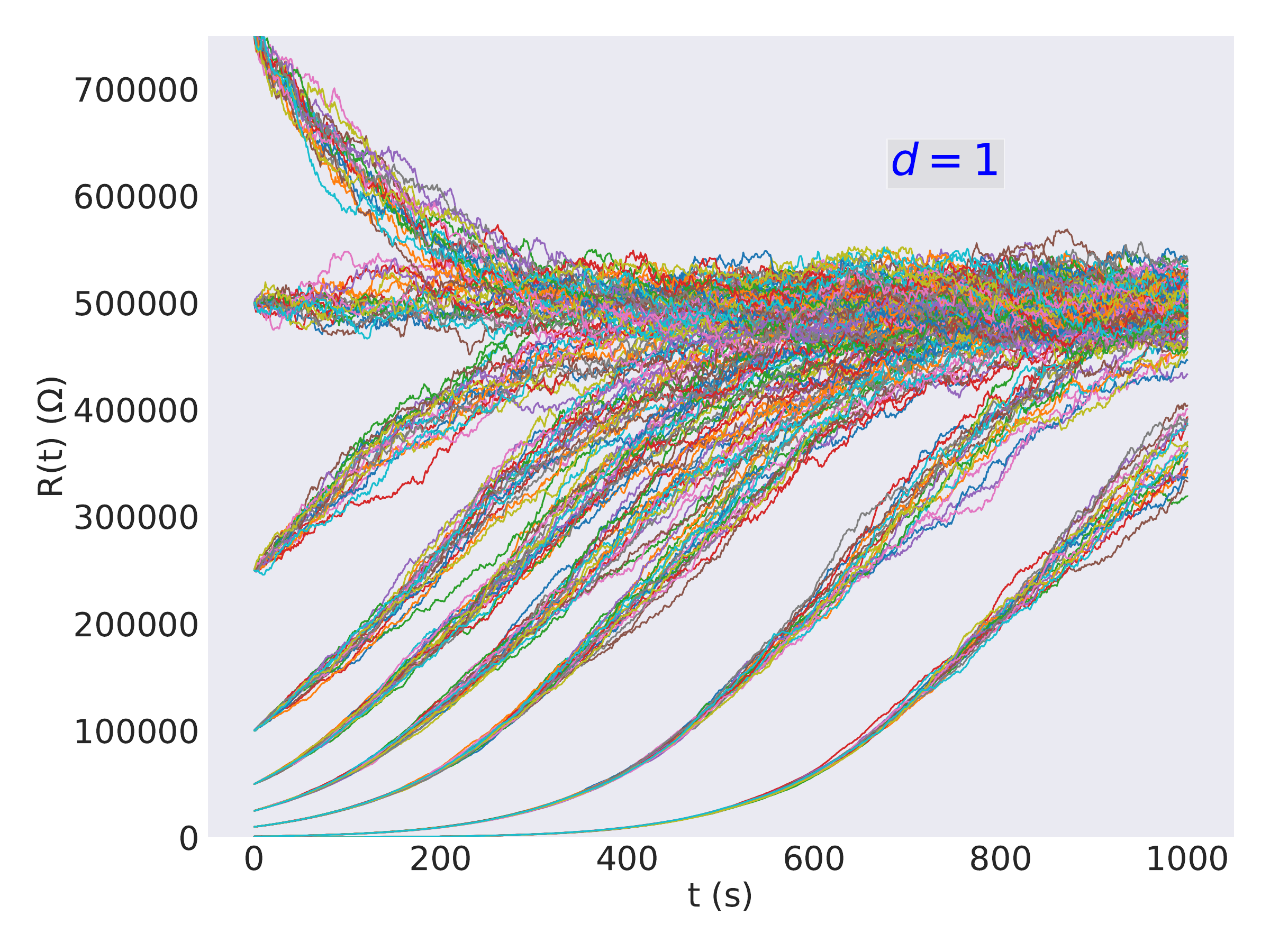}
    \label{fig:cgan_series_1}}
    \hfill
\subfloat[]{
    \includegraphics[width=0.3\textwidth]{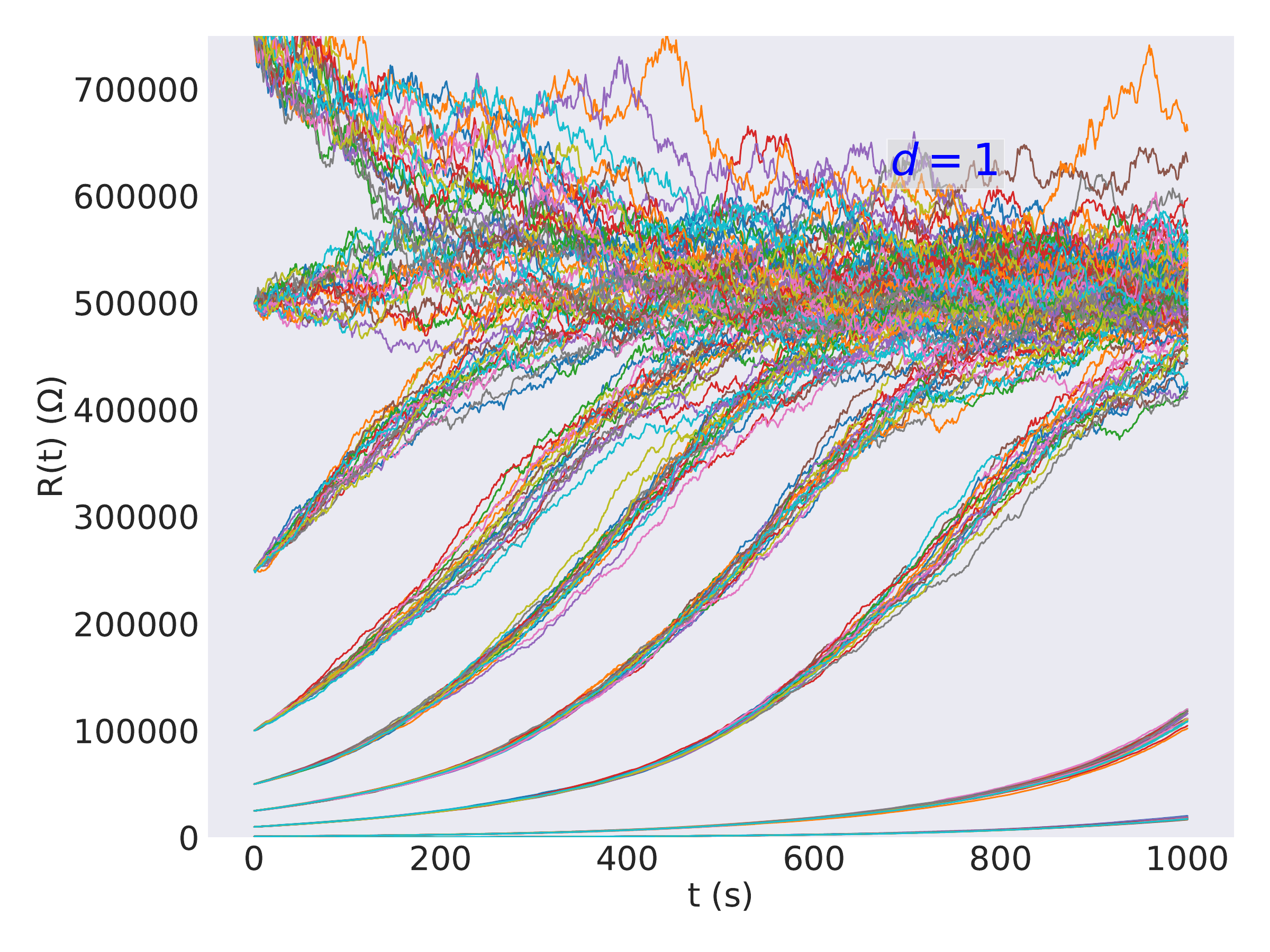}
    \label{fig:cgan_series_1_ablation}}
    \hfill
\subfloat[]{
    \includegraphics[width=0.3\textwidth]{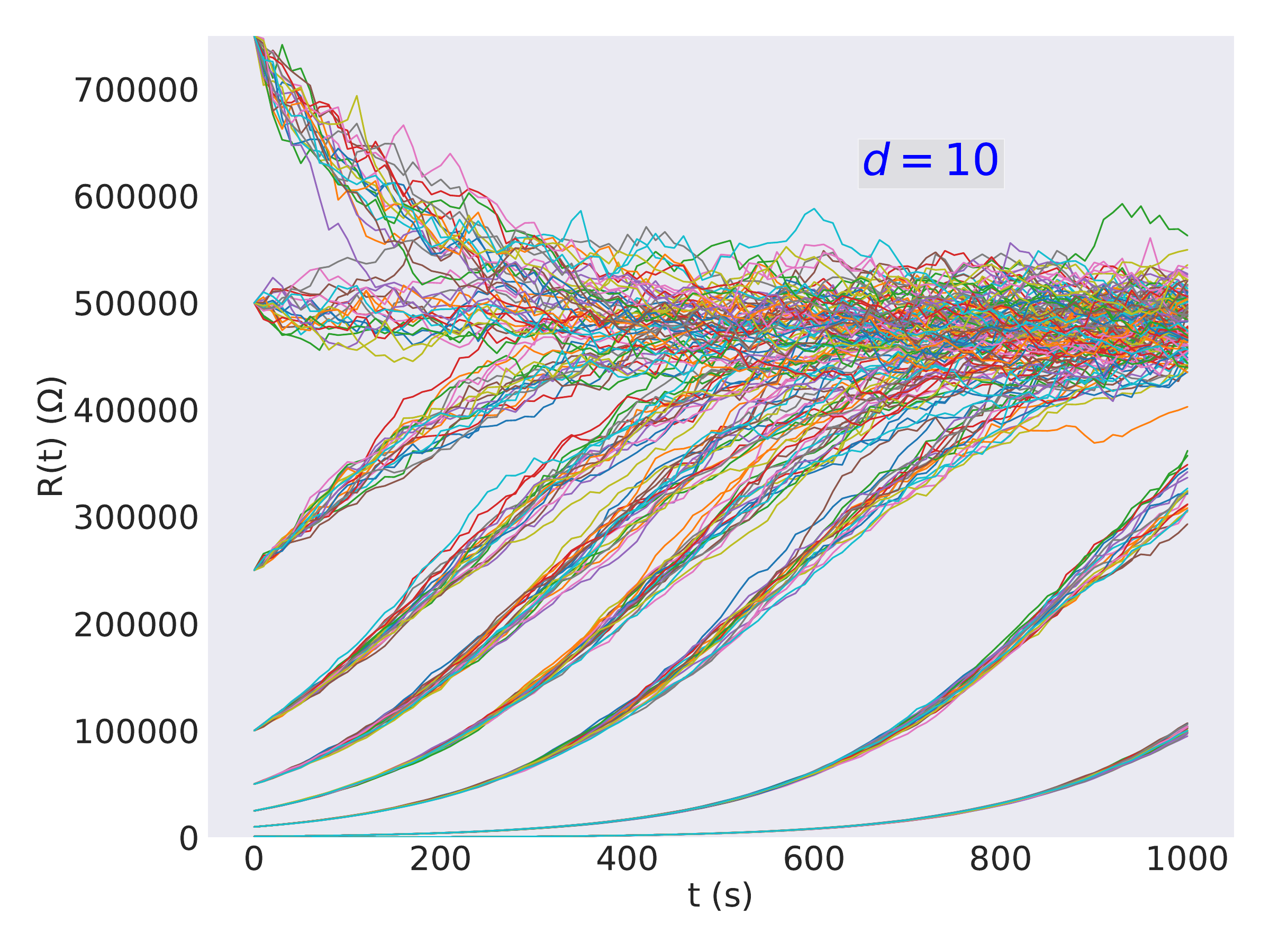}
    \label{fig:cgan_series_10}}
    \hspace*{\fill}
    \\
    \hspace*{\fill}
\subfloat[]{
    \includegraphics[width=0.3\textwidth]{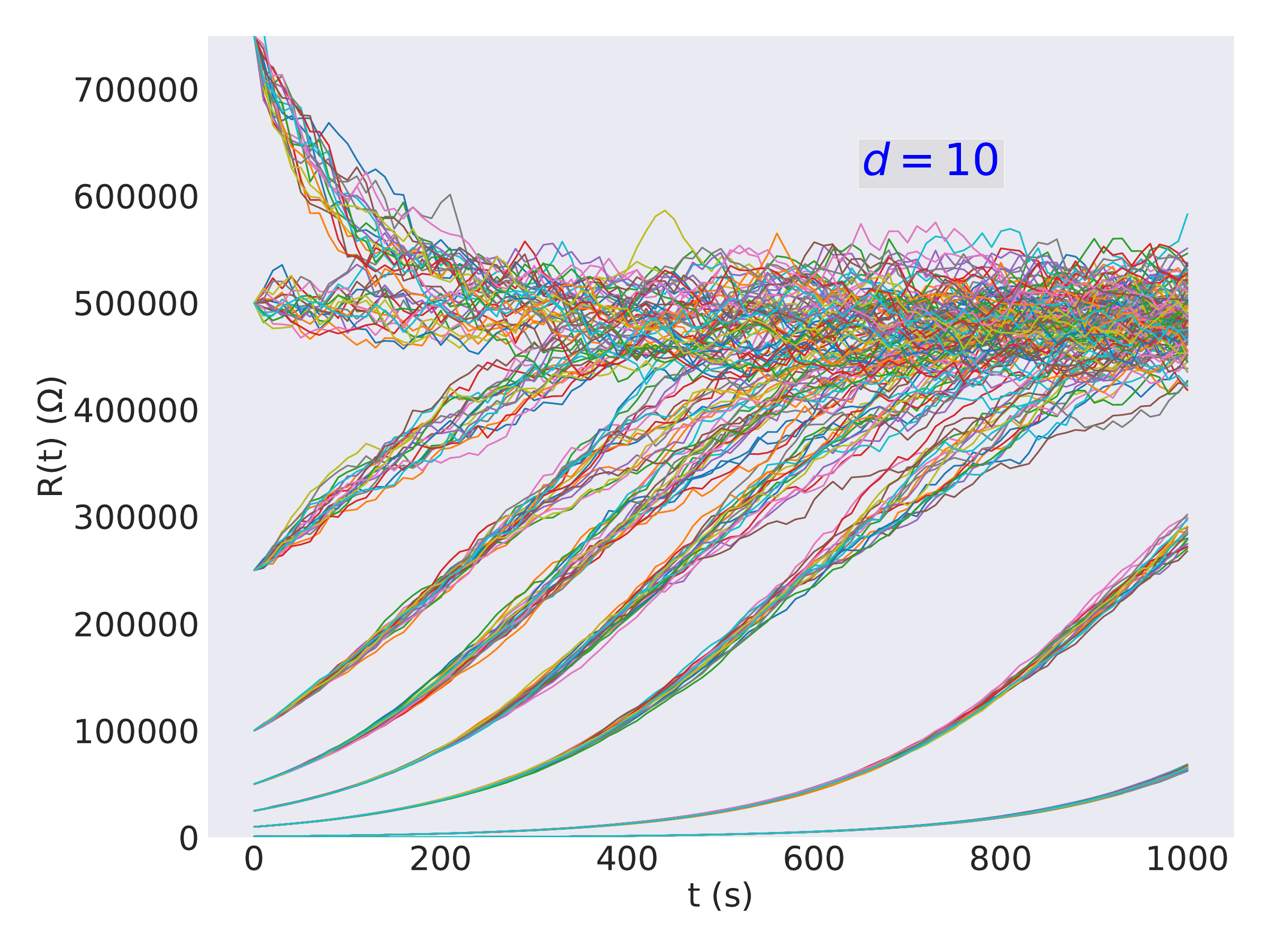}
    \label{fig:cgan_series_10_ablation}}
    \hfill
\subfloat[]{
    \includegraphics[width=0.3\textwidth]{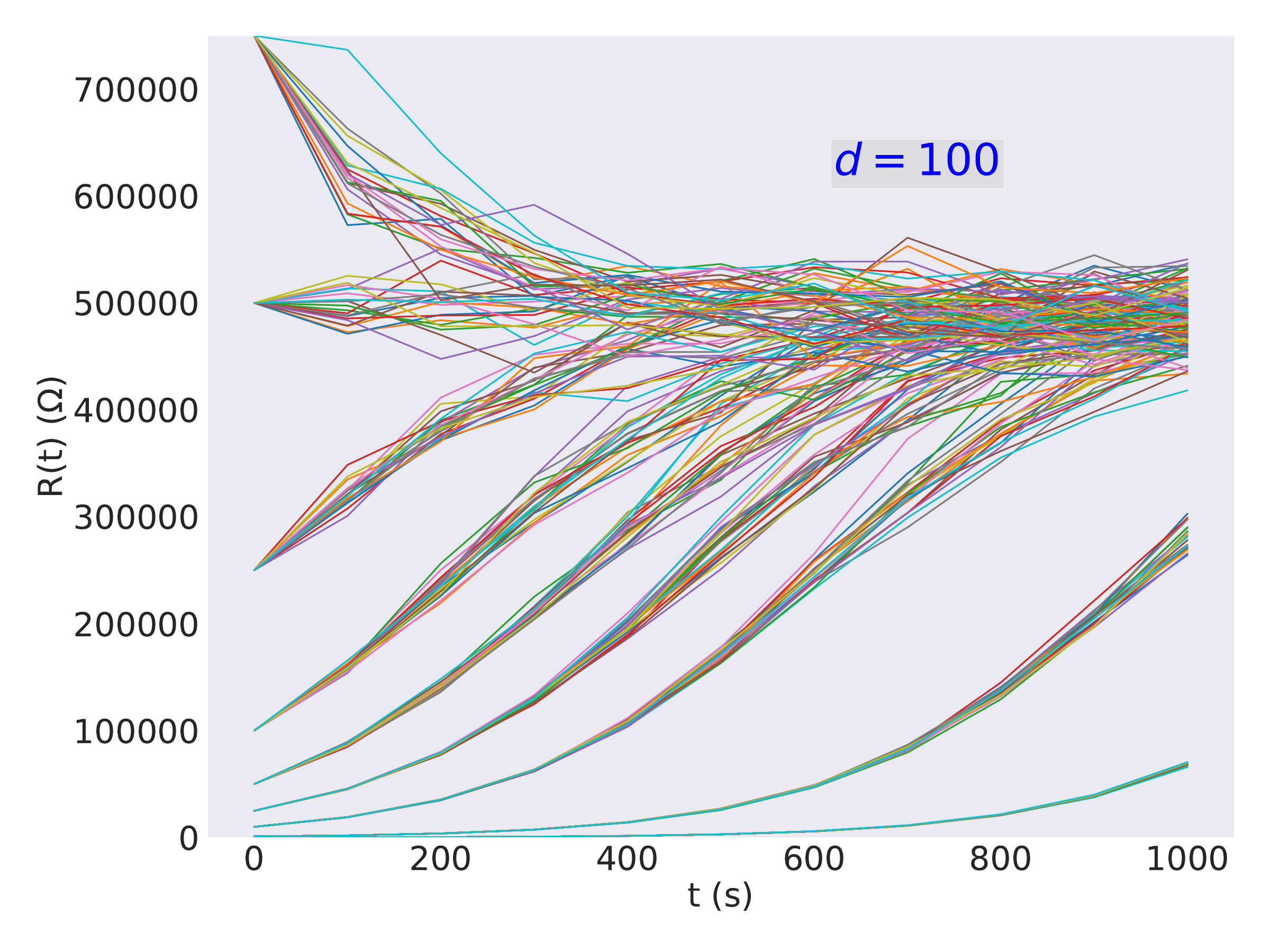}
    \label{fig:cgan_series_100}}
    \hfill
\subfloat[]{
    \includegraphics[width=0.3\textwidth]{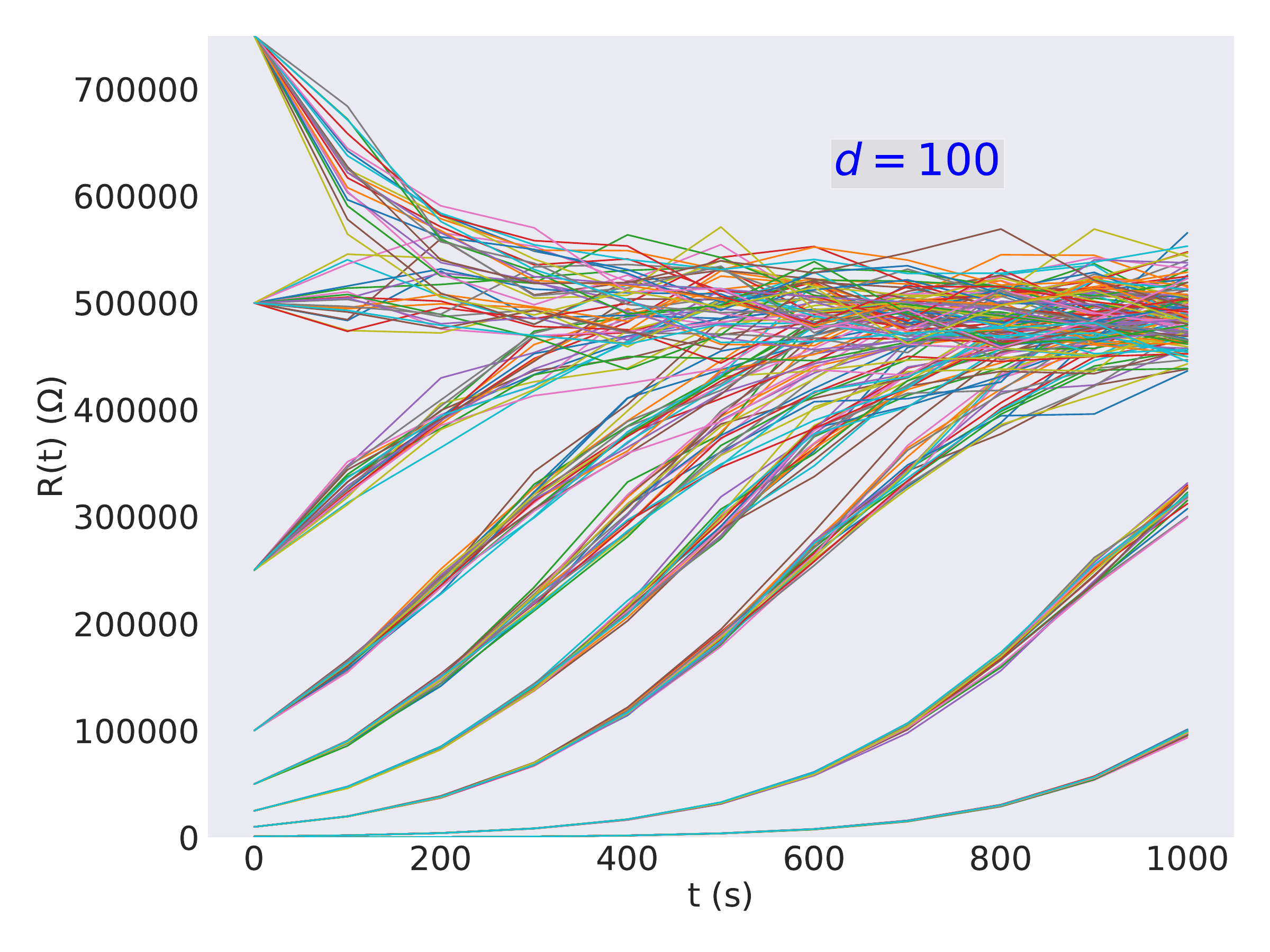}
    \label{fig:cgan_series_100_ablation}}
    \hspace*{\fill} \\
    \hspace*{\fill}
\subfloat[]{
    \includegraphics[width=0.3\textwidth]{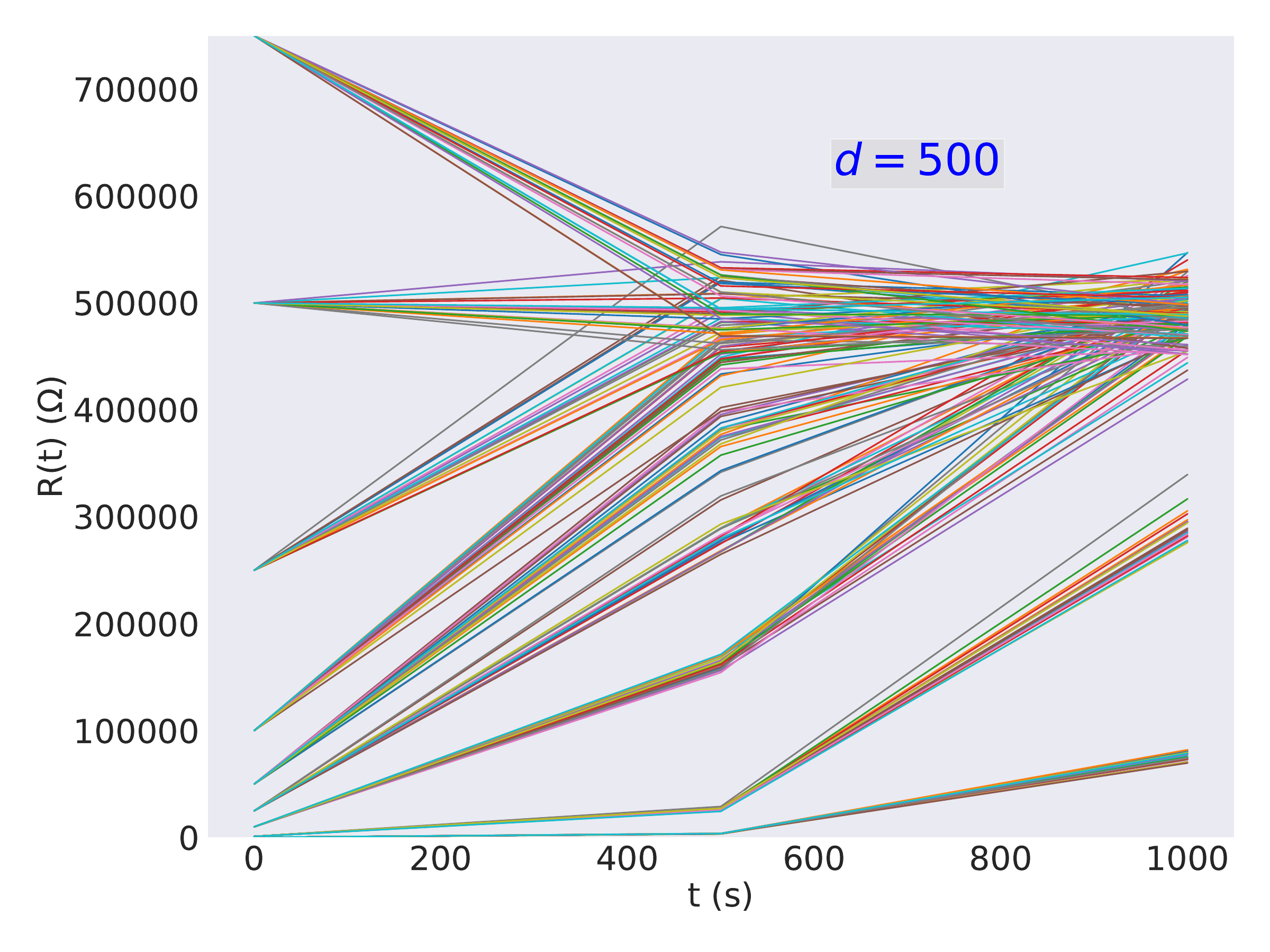}
    \label{fig:cgan_series_500}}
    \hfill
\subfloat[]{
    \includegraphics[width=0.3\textwidth]{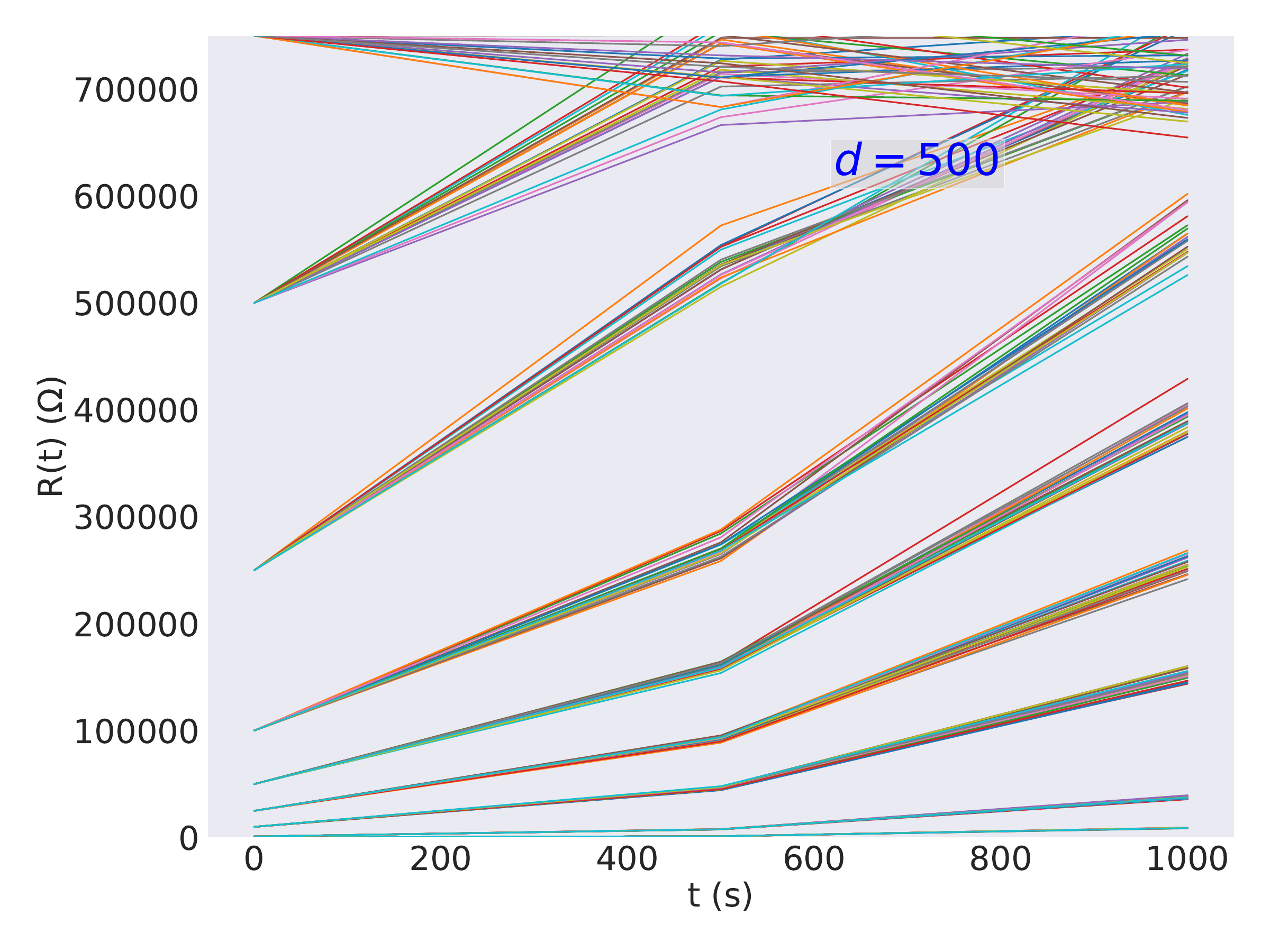}
    \label{fig:cgan_series_500_ablation}}
    \hspace*{\fill}
\caption{Generated series for different delays with (left figures \ref{fig:cgan_series_1}, \ref{fig:cgan_series_10}, \ref{fig:cgan_series_100}, \ref{fig:cgan_series_500}) and without (right figures \ref{fig:cgan_series_1_ablation}, \ref{fig:cgan_series_10_ablation}, \ref{fig:cgan_series_100_ablation}, \ref{fig:cgan_series_500_ablation}) the delay discriminator, for an equivalent delay of 1000, recurrently evaluated for different delay conditions \(d \in \{1, 10, 100, 500\}\).}
\label{fig:cgan_series}
\Description[Model generated series values for different delays.]{A series of plots showing the generative model's output for different delay conditions, with the model generating similar time series for different initial resistance values, regardless of the delay.}
\end{figure}

Below, we evaluate the model by recurrently generating time series data for a variety of delay conditions. This is done in order to demonstrate the model's ability to generate visually realistic series for a variety of initial resistance values and conditions, and to exhibit the lack of closed loop errors as part of the recurrent evaluation. We generate 20 series for each initial resistance value for a range of initial resistance values (for any given delay). In each case, we generate for a number of steps equal to \(\lceil (1000+d)/d \rceil\), where \(d\) is the delay condition. Example series generations are shown for the delay conditions of \(1, 10, 100, 500\) in Figure~\ref{fig:cgan_series}. These can be compared to the real series in the dataset, shown in Figure~\ref{fig:drift_data}, showing that the network has learnt to match the equilibrium point of convergence of 500\(k\Omega\) for a range of delay conditions and initial resistance values.

\subsection{Conditioned Statistics Evaluation}
\label{sec:cgan_generalisation}

\begin{figure}[h]
\centering
\hspace*{\fill}
\subfloat[Mean comparison.]{
    \includegraphics[width=0.4\textwidth]{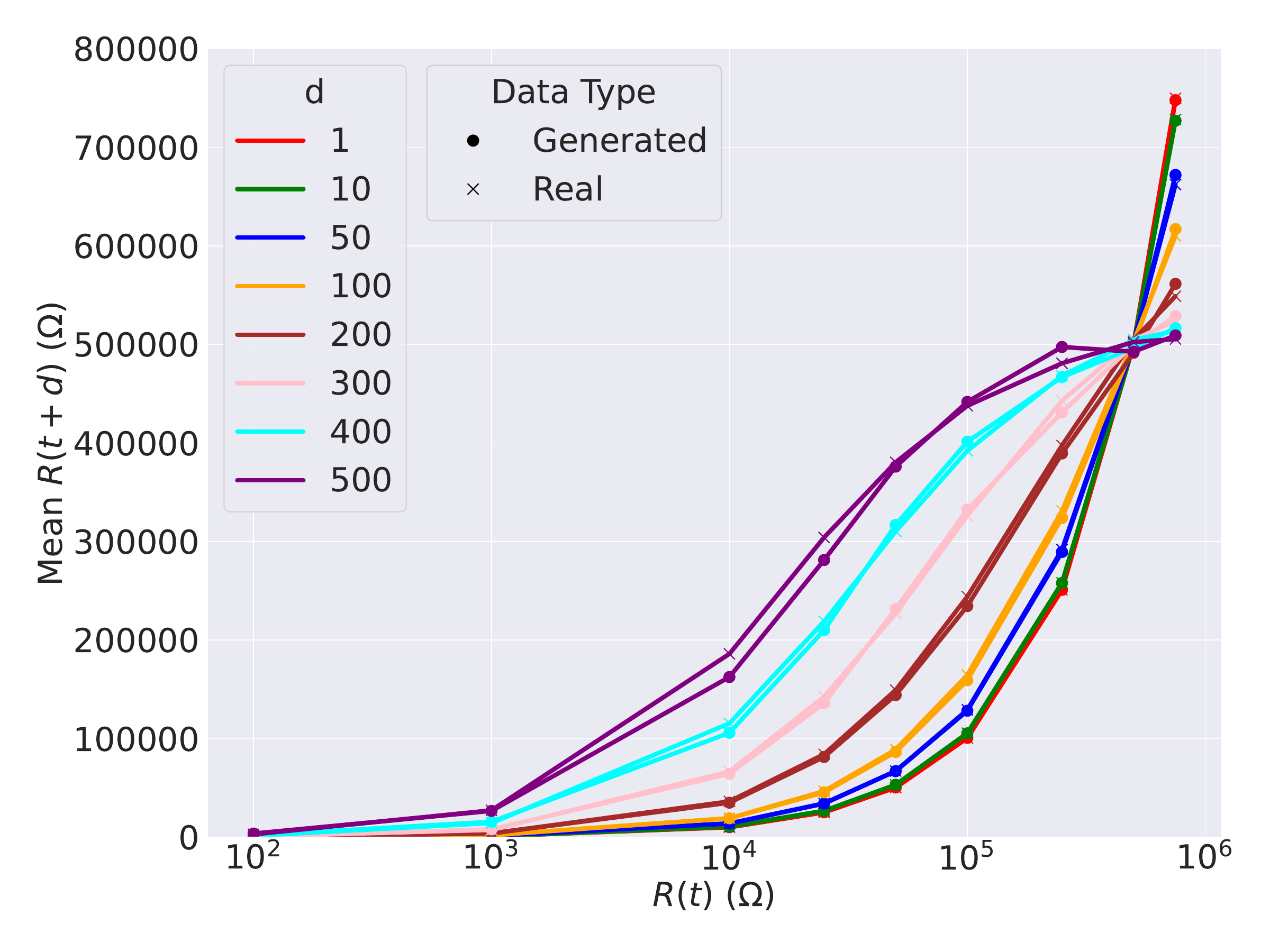}
    \label{fig:statistics_means_real}}
    \hfill
\subfloat[Standard deviation comparison.]{
    \includegraphics[width=0.4\textwidth]{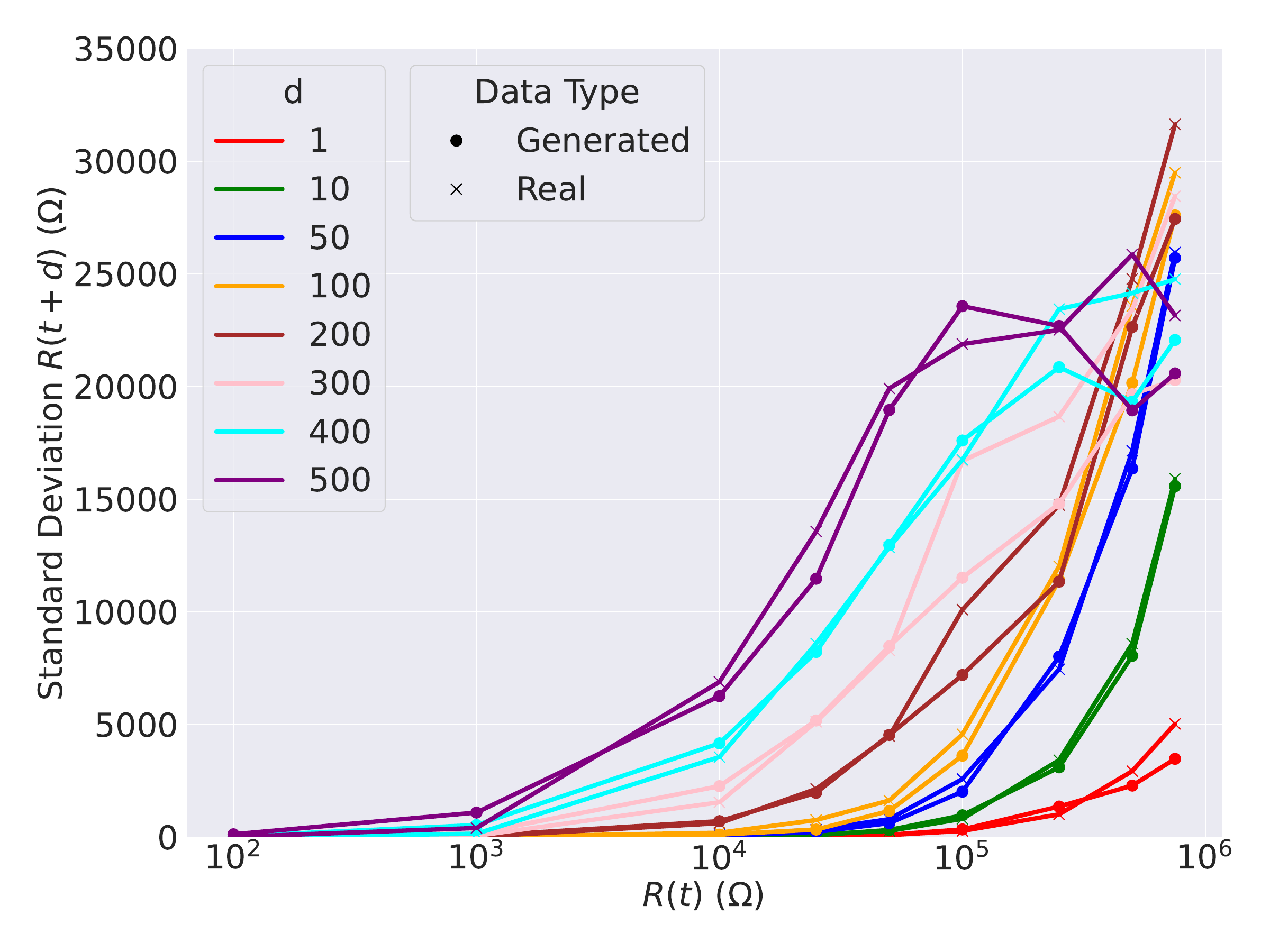}
    \label{fig:statistics_stds_real}}
    \hspace*{\fill} \\
    \hspace*{\fill}
\subfloat[Mean comparison for delay discriminator ablation experiment.]{
    \includegraphics[width=0.4\textwidth]{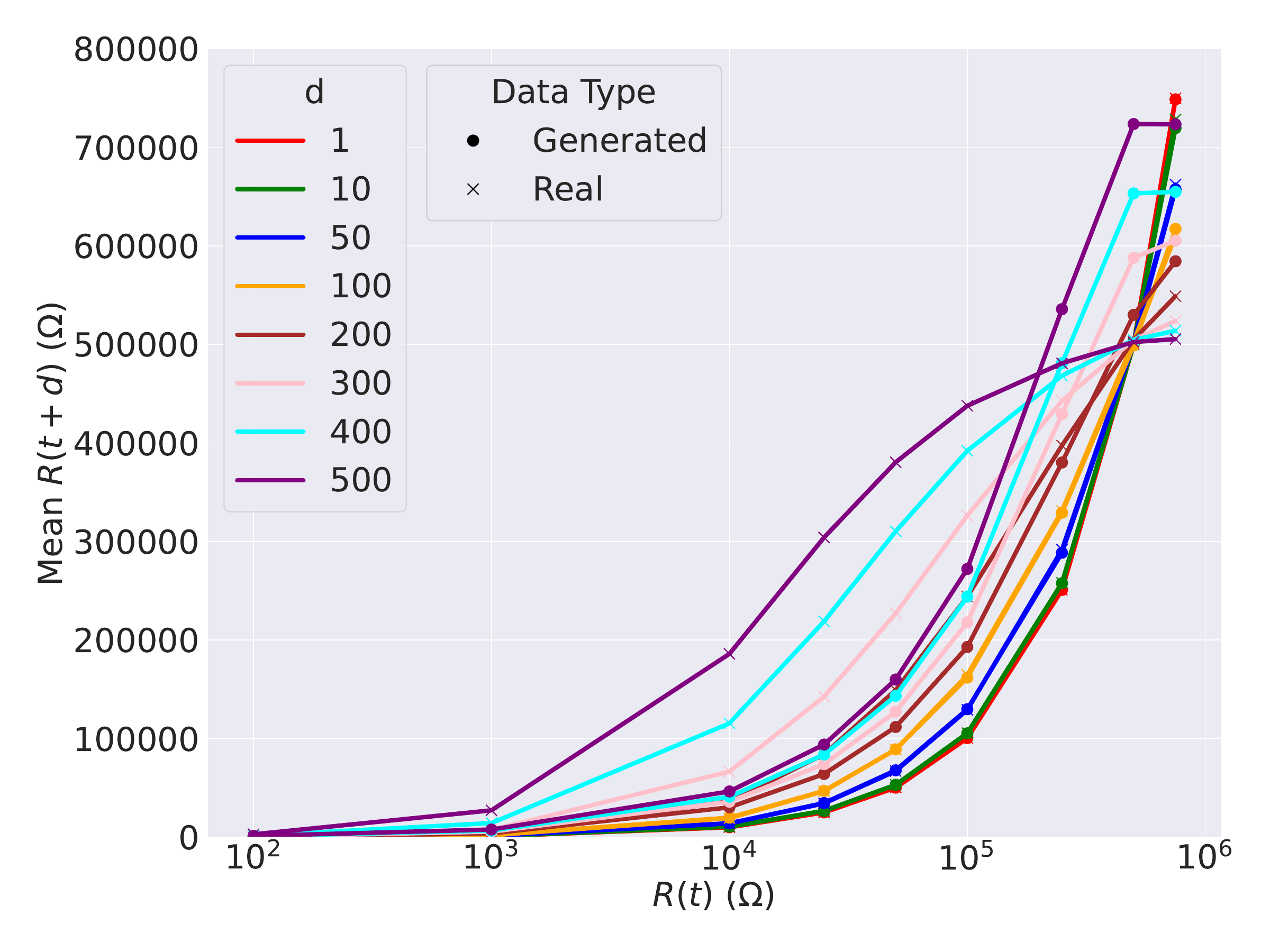}
    \label{fig:statistics_means_ablation}}
    \hfill
\subfloat[Standard deviation comparison for delay discriminator ablation experiment.]{
    \includegraphics[width=0.4\textwidth]{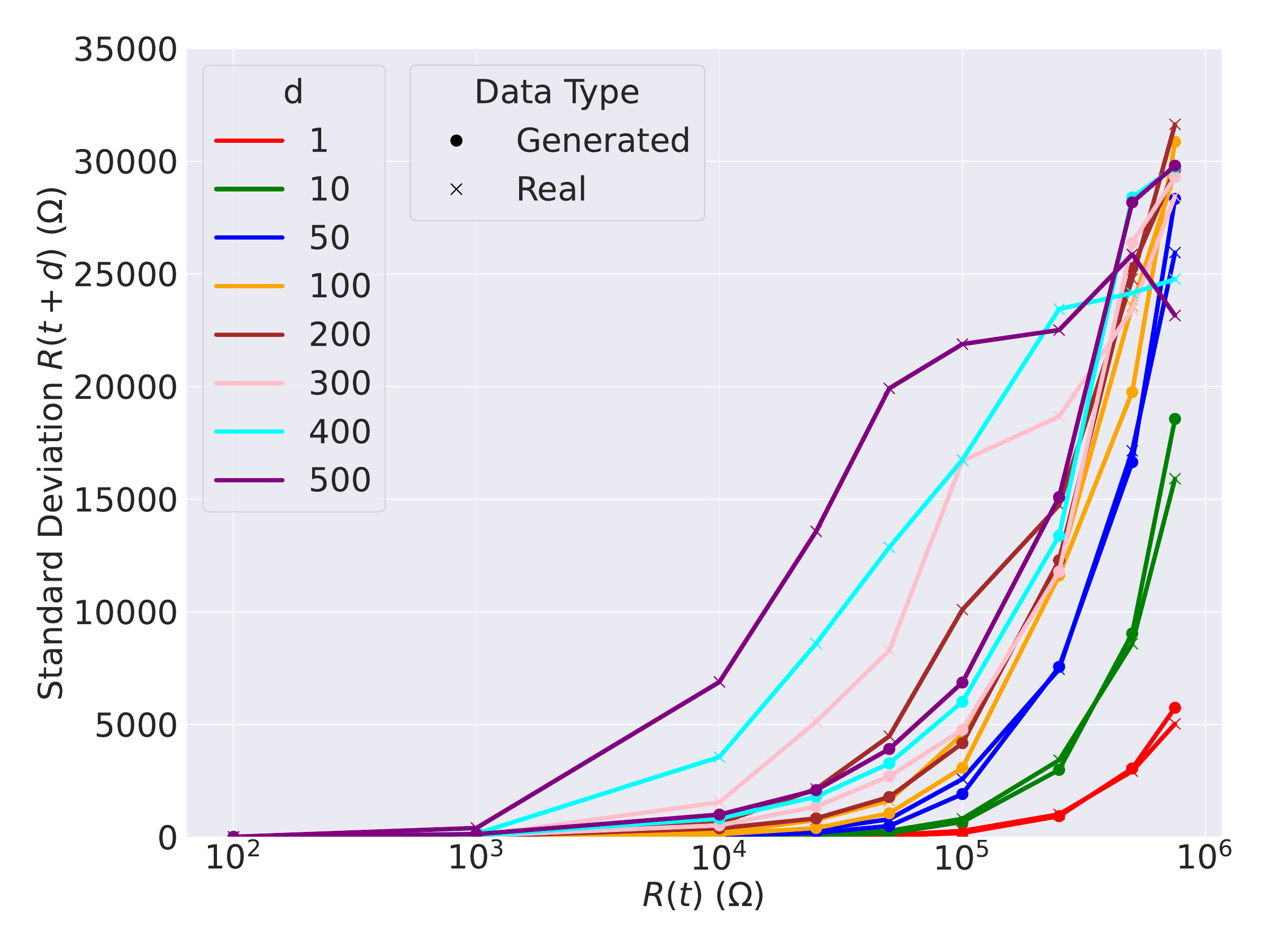}
    \label{fig:statistics_stds_ablation}}
    \hspace*{\fill}
\caption{Means and standard deviations of the final resistance conditioned on different combinations of initial resistances and delays, for the generator trained using the delay discriminator (Figure~\ref{fig:statistics_means_real}~and~\ref{fig:statistics_stds_real}) and trained without the delay discriminator as an ablation experiment (Figures~\ref{fig:statistics_means_ablation}~and~\ref{fig:statistics_stds_ablation}). The empirical statistics for each condition are plotted alongside those obtained from the ground truth model, in order to demonstrate the ability of the model to fit to the true distribution. We see a marked improvement in the statistical moment matching when using the delay discriminator. Legend entries denote the delay, \(d\) that the statistics are evaluated for, with `real' entries being the statistics derived from the ground truth event-based model.}
\label{fig:statistics}
\Description[Statistics of the modelling approach.]{The means and standard deviations of the model output for different initial resistance and delay values, compared to the ground truth model. The same statistics are also shown for a delay discriminator ablation experiment.}
\end{figure}

In Figure~\ref{fig:statistics}, we present a comparison of the statistical means and standard deviations of the learned conditional distribution to those estimated from the real model. In order to estimate these statistics, we generate 100 samples of the \ac{cGAN} model output for each set of conditions, and do the same for the ground truth event-based model (also using 100 samples per parameter estimate). We plot comparisons for both the proposed \ac{cGAN} model, and for an ablation experiment excluding the delay discriminator.

\begin{figure}[h]
\centering
    \hspace*{\fill}
\subfloat[Ground truth model.]{
    \includegraphics[width=0.3\textwidth]{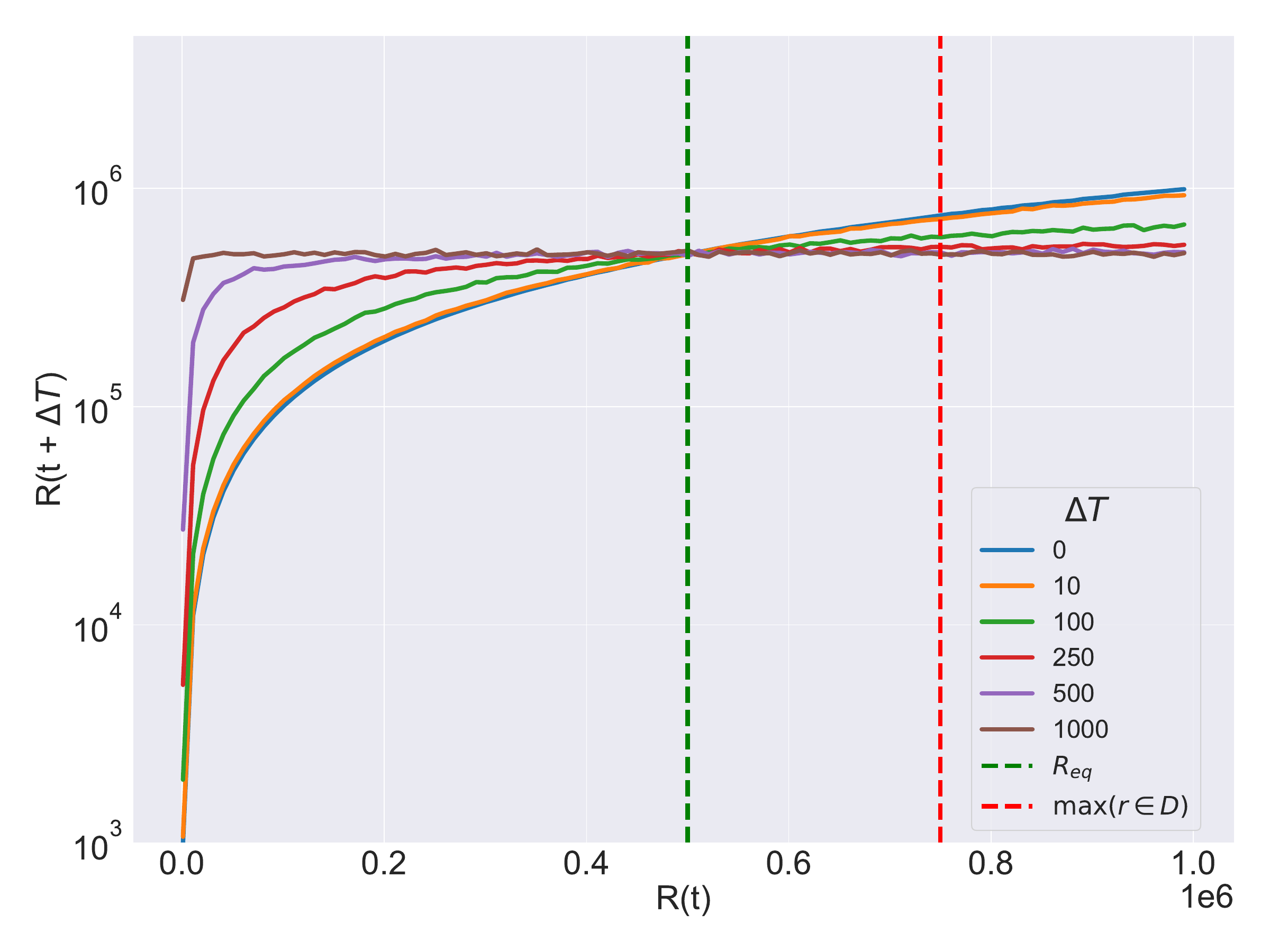}
    \label{fig:pairs_conditioned_real}}
    \hfill
\subfloat[Conditional GAN.]{
    \includegraphics[width=0.3\textwidth]{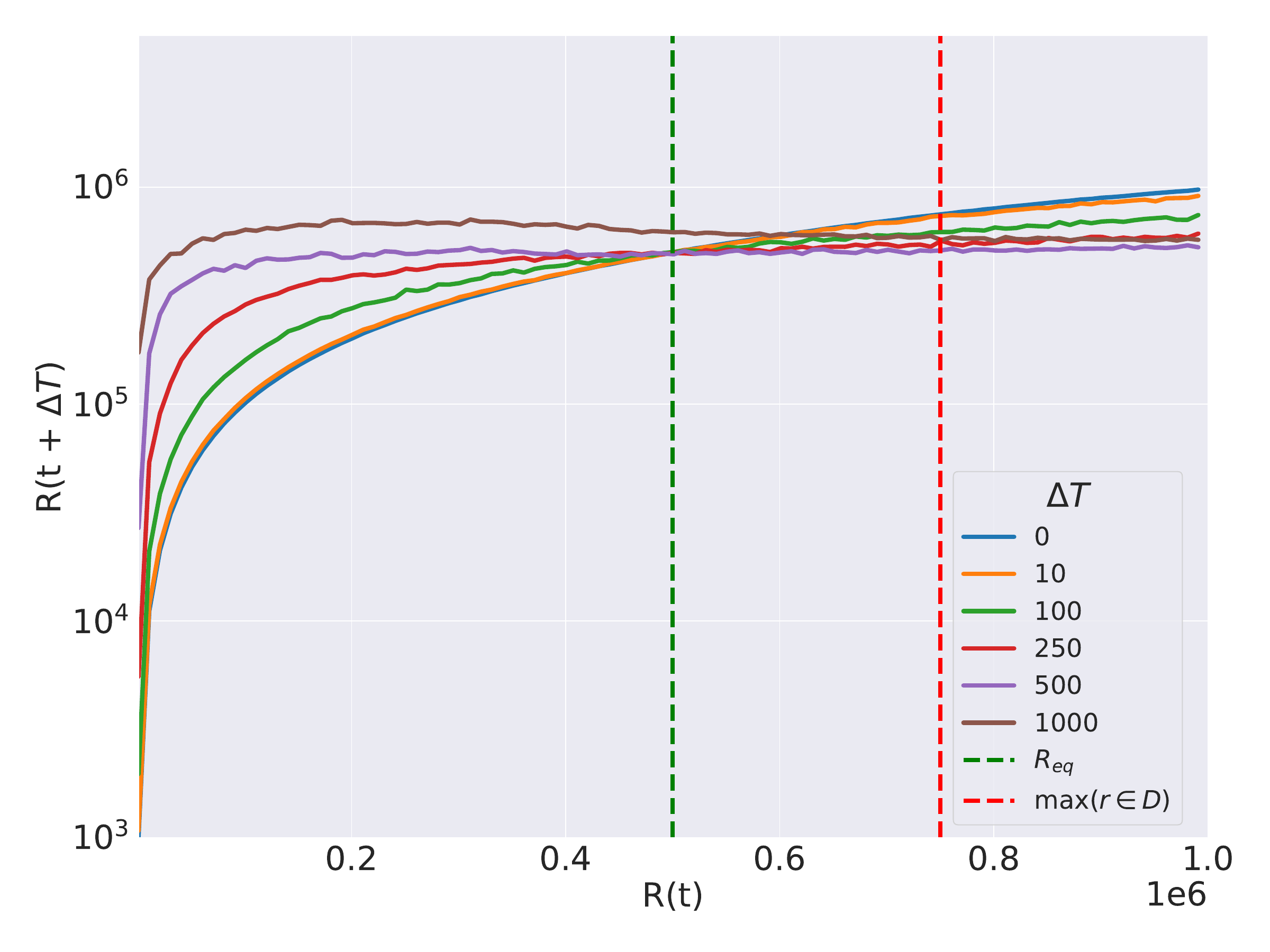}
    \label{fig:pairs_conditioned_nonegative}}
    \hfill
\subfloat[Conditional GAN with recurrent evaluation.]{
    \includegraphics[width=0.3\textwidth]{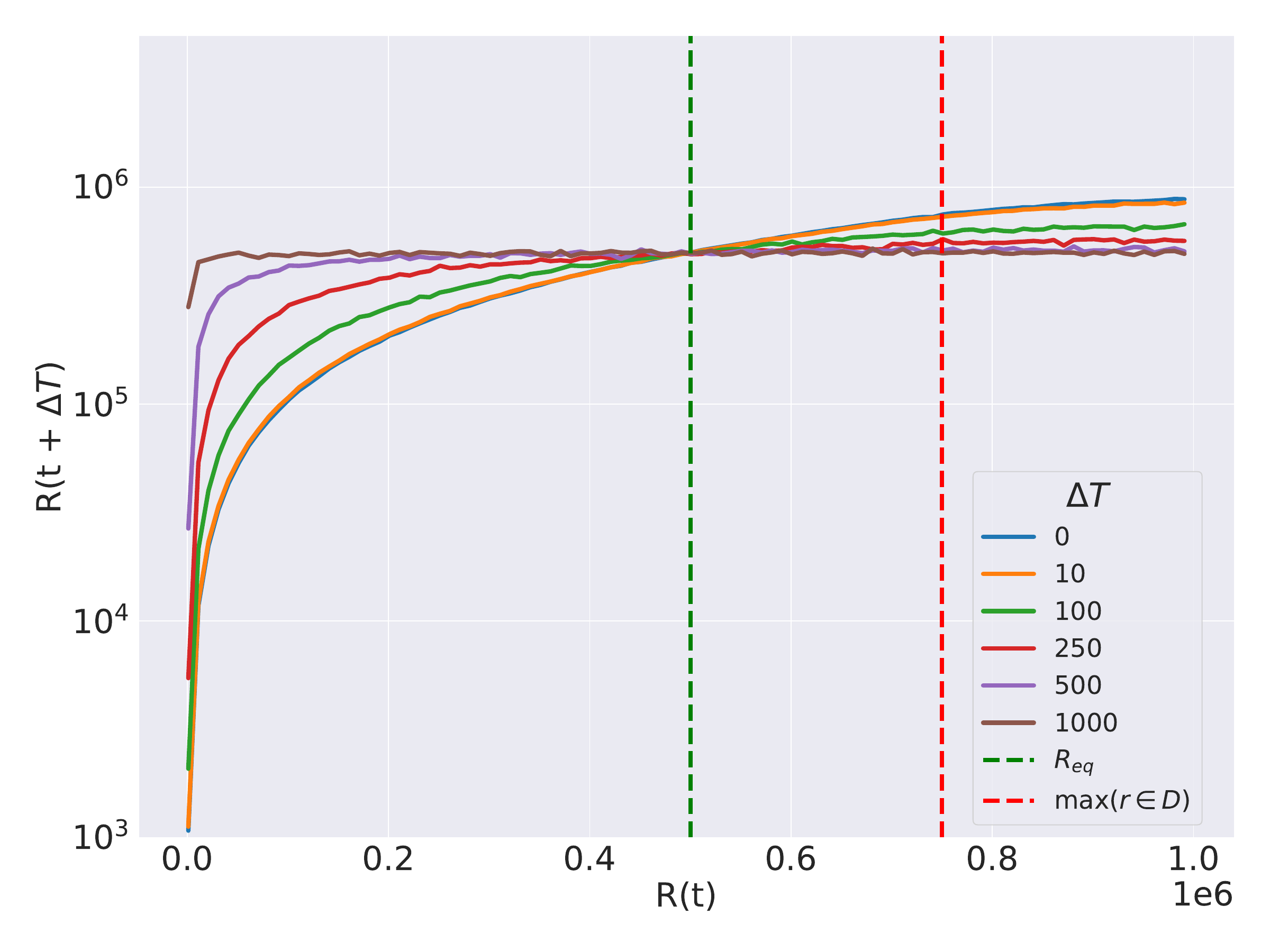}
    \label{fig:pairs_conditioned_nonegative_split}}
    \hspace*{\fill}
    \\
\centering
    \hspace*{\fill}
\subfloat[Delay discriminator ablation.]{
    \includegraphics[width=0.3\textwidth]{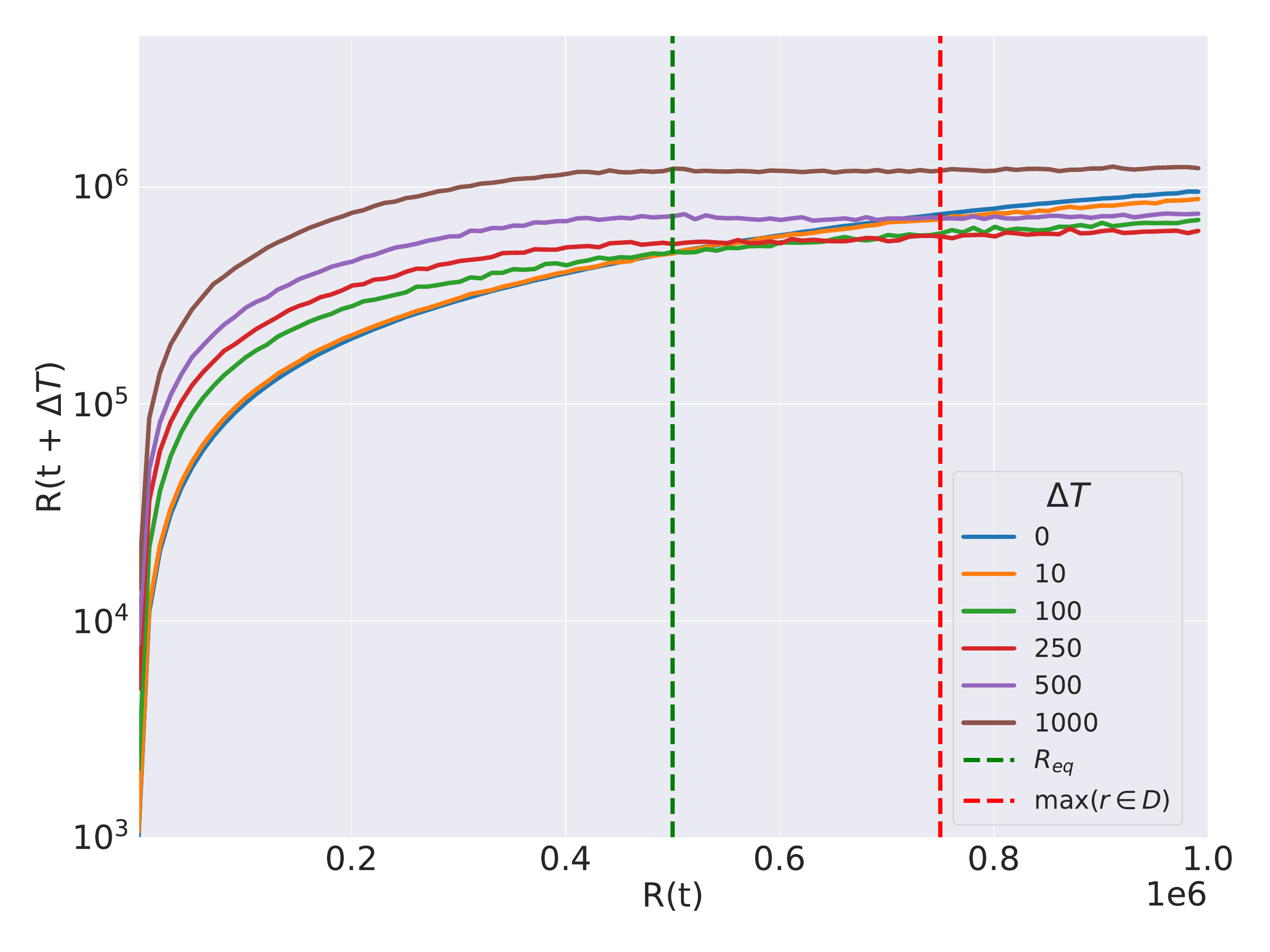}
    \label{fig:pairs_conditioned_ddablation}}
    \hfill
\subfloat[Delay discriminator ablation with recurrent evaluation.]{
    \includegraphics[width=0.3\textwidth]{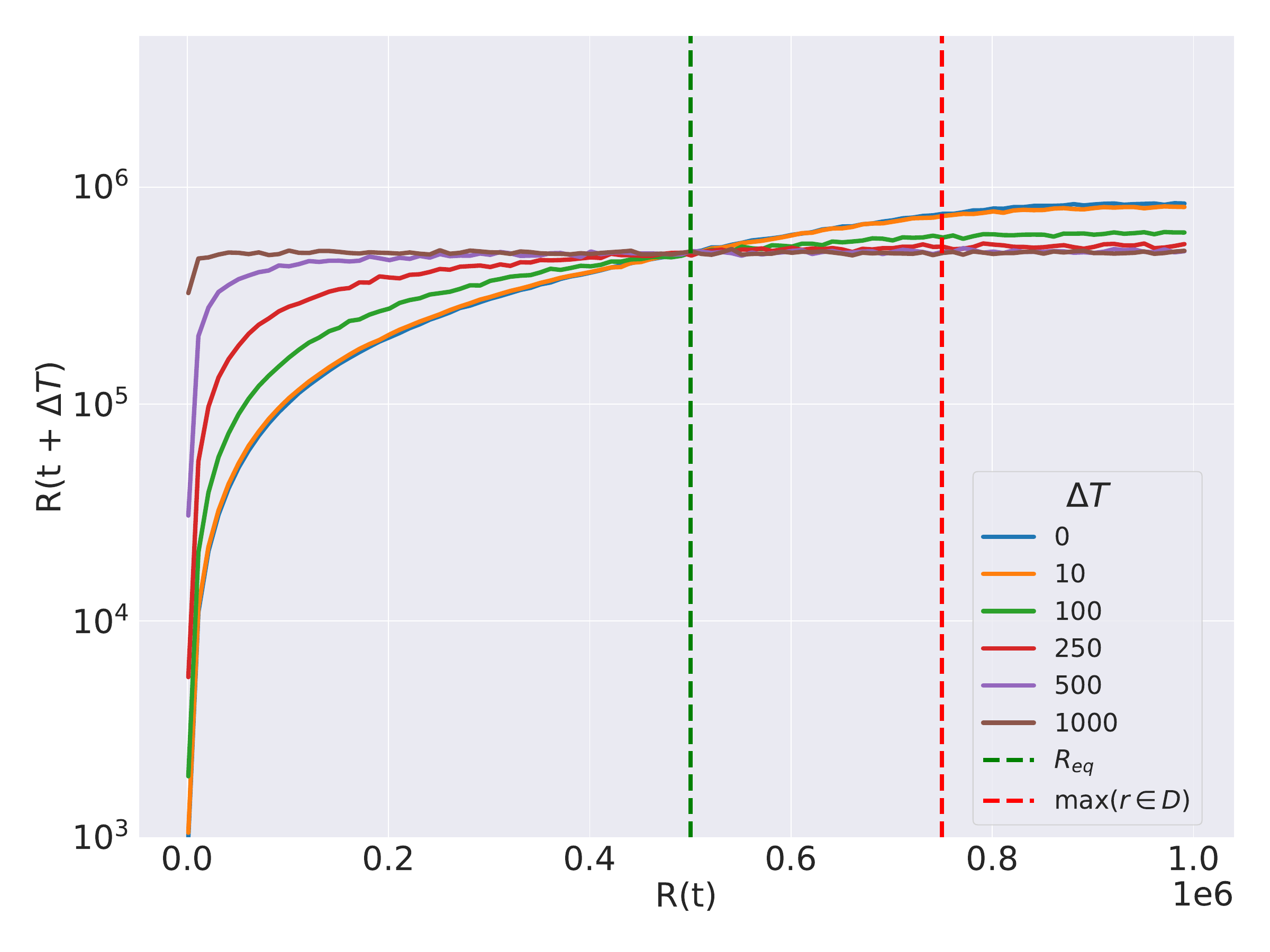}
    \label{fig:pairs_conditioned_ddablation_split}}
    \hspace*{\fill}
\caption{Empirically sampled final resistance values for given initial resistance and delay values. Figures~\ref{fig:pairs_conditioned_real} and \ref{fig:pairs_conditioned_nonegative} show the statistics for the ground truth event-based model and \ac{cGAN}, respectively. Figure~\ref{fig:pairs_conditioned_nonegative_split} shows statistics for \ac{cGAN} evaluated recurrently with a delay of \(d/10\) and 10 recurrent steps. Figure~\ref{fig:pairs_conditioned_ddablation} shows an ablation experiment run without the delay discriminator, and Figure~\ref{fig:pairs_conditioned_ddablation_split} shows a recurrent evaluation of the delay discriminator ablation experiment with a delay of \(d/10\) and 10 recurrent steps.}
\label{fig:pairs_comparison}
\Description[Statistics of the conditioned final resistances of the generative model.]{Empirical statistics of the generative model, alongside an ablation experiment and the ground truth model statistics, for both a single step and recurrent evaluation approach.}
\end{figure}

Figure~\ref{fig:pairs_conditioned_real} shows final resistance values for given initial resistance and delay conditions, simulated using the true event-based model (the ground truth). Figure~\ref{fig:pairs_conditioned_nonegative} shows a comparison of the final resistance values for the same conditions, with simulations performed by the \ac{cGAN} model. For delays, \(\Delta T\), up to \(\Delta T = 500s\), which was the maximum value of \(d_{dd}\) (the delay discriminator delay), we see that the model's performance is accurate. Although, for delays above this (\(1000s\)) the \ac{cGAN} model's similarity to the true statistics drops.

Figure~\ref{fig:pairs_conditioned_nonegative_split} shows the model output for the same conditions, but using a recurrent evaluation with a delay of \(d = \Delta T/10\), for 10 closed loop recursions (feeding the output of the model back to itself as input). We see that recurrently evaluating the model with a smaller delay enables more accurate recreation of the statistics for this larger total delay.

We see, in Figure~\ref{fig:pairs_conditioned_ddablation}, the model output for the same conditions, for an ablation experiment run without the delay discriminator.
The same recurrent evaluation setting for the delay discriminator is shown in Figure~\ref{fig:pairs_conditioned_ddablation_split}. As can be seen, the model's accuracy in matching the statistics of the dataset for delays larger than those in the training data (i.e., \(d_{\text{max}} = 90s\)) is significantly degraded. When the generator is evaluated recurrently, we see an improved performance, due to the presence of these smaller delay conditions in the training data. The fact that more accurate modelling for larger delays requires recurrent evaluation using smaller delays in the absence of delay discrimination, justifies the use of delay discrimination in enabling computationally simple conditional distribution modelling for larger delays.

\section{End-to-End Quantisation Level Optimisation}
\label{sec:quantisation_application}
In order to demonstrate the advantages of our proposed generative modelling approach in the context of end-to-end learning, we here present an application of the differentiable model for the optimisation of quantisation levels and decoding boundaries for multilevel data storage on memristors. Here, the differentiable nature of the model is crucial in enabling optimisation of quantisation parameters using gradient descent. The computational efficiency in simulating outputs for large delay conditions - without the need for recurrent evaluation - enables fast Monte Carlo estimation of statistical parameters of interest, which can be used to improve the quantisation parameters using accurate error estimates derived from large numbers of samples, without problems associated with vanishing gradients.

\subsection{Problem Statement}

We consider the problem of storing discrete values on a noisy memristor, making use of the \ac{cGAN} model as a differentiable storage model that can be used for optimisation of the quantisation levels and boundaries, providing an automated method for selection of a storage and threshold decoding approach.

Let \(\mathcal{L}\) denote an ordered set of quantisation levels, and \(\mathcal{B}\), where \(B[0] \triangleq -\infty\) and \(B[|B|] \triangleq \infty\), denote an ordered set of quantisation boundaries. Note that the decoding boundaries for the \(i\)'th quantisation level, \(l_i\), which border the bin, are defined as the half-open interval \([\mathcal{B}[i], \mathcal{B}[i+1])\).
We consider the problem of maximising the number of quantisation levels, \(|\mathcal{L}|\), that can be decoded given a maximum tolerable threshold on the probability of error, \(\epsilon\), based on the decoding approach defined by \(\mathcal{B}\):
\begin{align}
    \max_{\mathcal{L}, \mathcal{B}} |\mathcal{L}| ~~~ \text{   s.t. } \frac{1}{|\mathcal{L}|}\sum_{l_i \in \mathcal{L}} \mathbb{E}\left[ \mathbf{1}_{[\mathcal{B}[i], \mathcal{B}[i+1])}\left( G(l_i, d) \right) \right] \leq \epsilon
    \label{eq:quantisation_problem_definition}
\end{align}
where \(d\) is the chosen delay, \(\mathbf{1}_A\) is the indicator function, and \(G(r, d)\) is the delay and initial resistance conditional resistive drift distribution. Note that here, we have formulated the problem using the \ac{cGAN} generator as a variational approximation of the true conditional distribution in order to make the problem differentiable and optimisable through gradient descent, as well as enabling quick Monte Carlo estimation of the average decoding error.

\subsection{Method}

We use a stochastic gradient descent approach to optimise the quantisation levels and decoding boundaries. Our loss consists of three terms:

\begin{enumerate}
    \item \textbf{Crossover loss}: Monte-Carlo estimate of integral of the probability density falling outside the given interval, for each distribution marginalised over each of the levels \(l_i \in \mathcal{L}\) as an initial resistance value.
    \item \textbf{Ordering regularisation}: We require that the order of the quantisation bins and their respective boundaries be strictly preserved through penalising any violation of the ordering with a squared error. We enforce the requirement that quantisation levels must lie within their decoding boundaries.
    \item \textbf{Dynamic range regularisation}: We specify an allowable range - \([r_{\text{qmin}}, r_{\text{qmax}}]\) - for the quantisation levels, ensuring that the upper and lower quantisation levels do not fall outside the boundaries of this range, through a squared error penalty.
\end{enumerate}

We note that the loss function as given is non-zero only if there is an error in the decoding process, or a boundary violation. In the absence of such conditions, the gradient with respect to the quantisation levels and decoding boundaries will be zero. In order to avoid the problem of zero gradients for the regularisation terms when ordering and dynamic range constraints are satisfied, or when the probability of error is low, we use a margin, \(\rho\), such that the loss is non-zero close to points of constraint violation.

Let \(h(x, y) \triangleq (\min (x-y, 0))^2\). We can define our loss function for a given delay, \(L_d\), as follows for the margin:
\begin{align}
    L_d(\mathcal{L}, \mathcal{B}) &:= \sum_{l_i \in \mathcal{L}} \mathbb{E}\left[h(\mathcal{B}[i] + \rho, G(l_i, d)) + h(G(l_i, d) + \rho, \mathcal{B}[i+1])\right] \nonumber \\
    &+ \lambda_1 \sum_{l_i \in \mathcal{L}} h(\mathcal{B}[i] + \rho, l_i) + h(l_i + \rho, \mathcal{B}[i+1]) \nonumber \\
    &+ \lambda_2 (h(r_{\text{qmin}} + \rho, l_0) + h(l_0 + \rho, r_{\text{qmax}}))
    \label{eq:loss_function_quantisation}
\end{align}
where \(\lambda_1\) and \(\lambda_2\) are regularisation factors.

We use an Adam optimiser \cite{kingmaAdamMethodStochastic2017} with an initial learning rate of \(1\times10^{-3}\) and a learning rate schedule that reduces the learning rate by a factor of \(0.9\) whenever the empirical loss does not improve for more than 25 optimisation steps. For each step, we estimated the expected value in Equation~\ref{eq:loss_function_quantisation} through a Monte Carlo estimate over 32 trials. We run our algorithm for each delay value and number of quantisation levels until convergence is observed, or for \(10,000\) steps, whichever is sooner. We set the allowable input range to \([10000, 500000]\). We set \(\lambda_1 = 10\) and \(\lambda_2 = |\mathcal{L}|/2\) to discourage violation of the constraints, and set \(\rho = 0.05\).

\subsection{Results}

\begin{figure}
    \centering
    \hspace*{\fill}
    \subfloat[]{\includegraphics[width=0.45\linewidth]{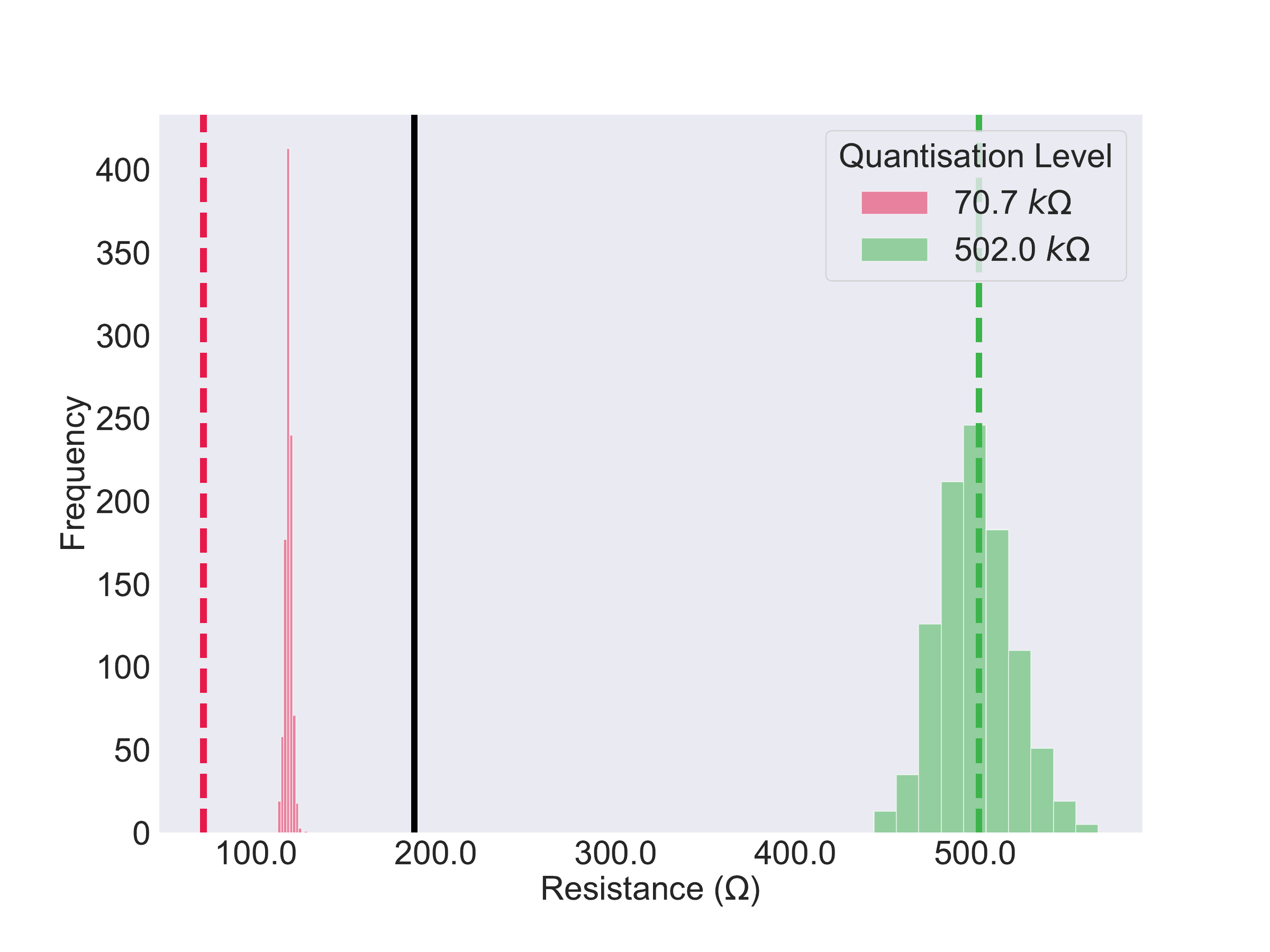}
    \label{fig:quantisation_2}}
    \hfill
    \subfloat[]{\includegraphics[width=0.45\linewidth]{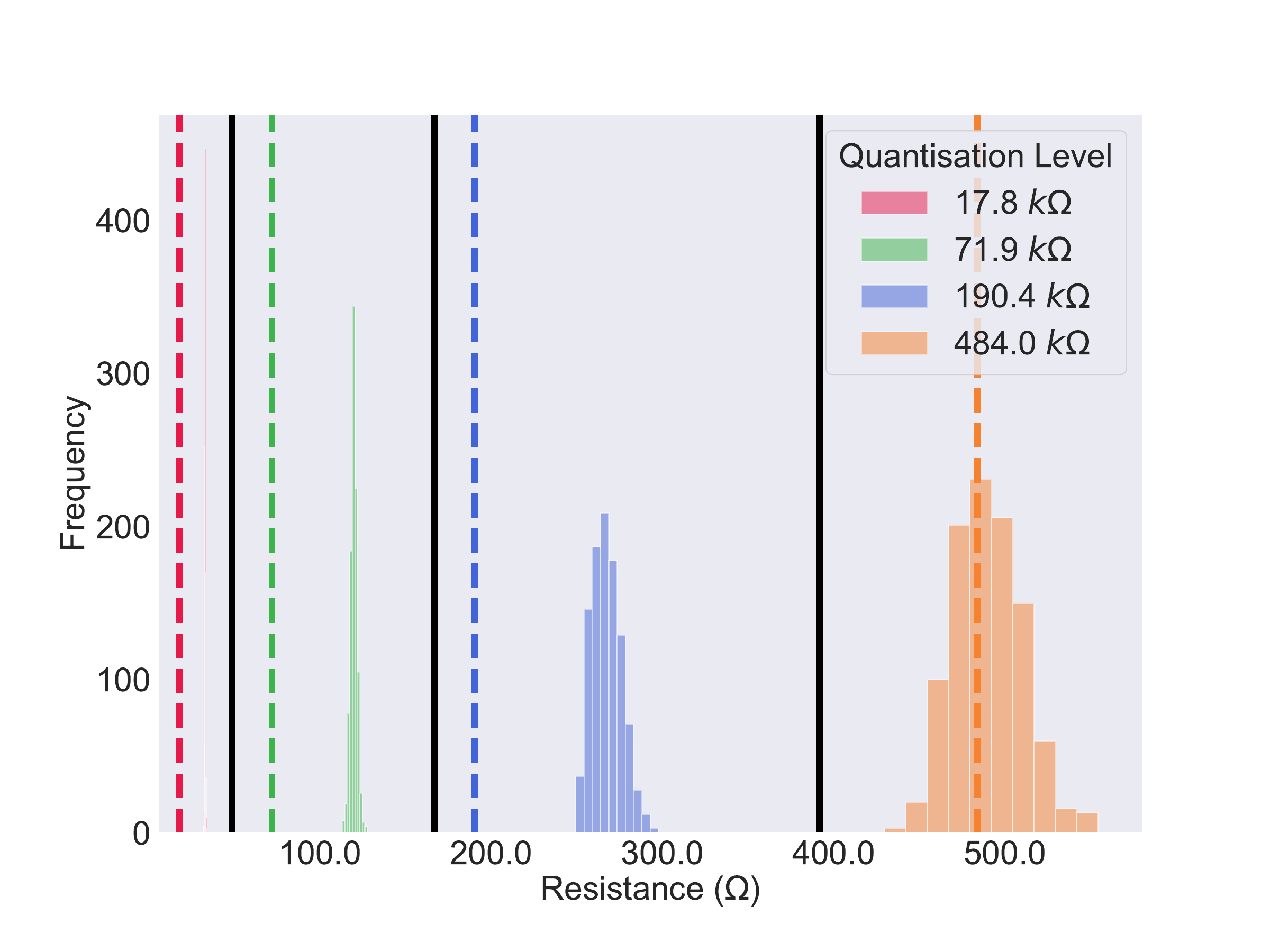}
    \label{fig:quantisation_4}}
    \hspace*{\fill} \\
    \hspace*{\fill}
    \subfloat[]{\includegraphics[width=0.45\linewidth]{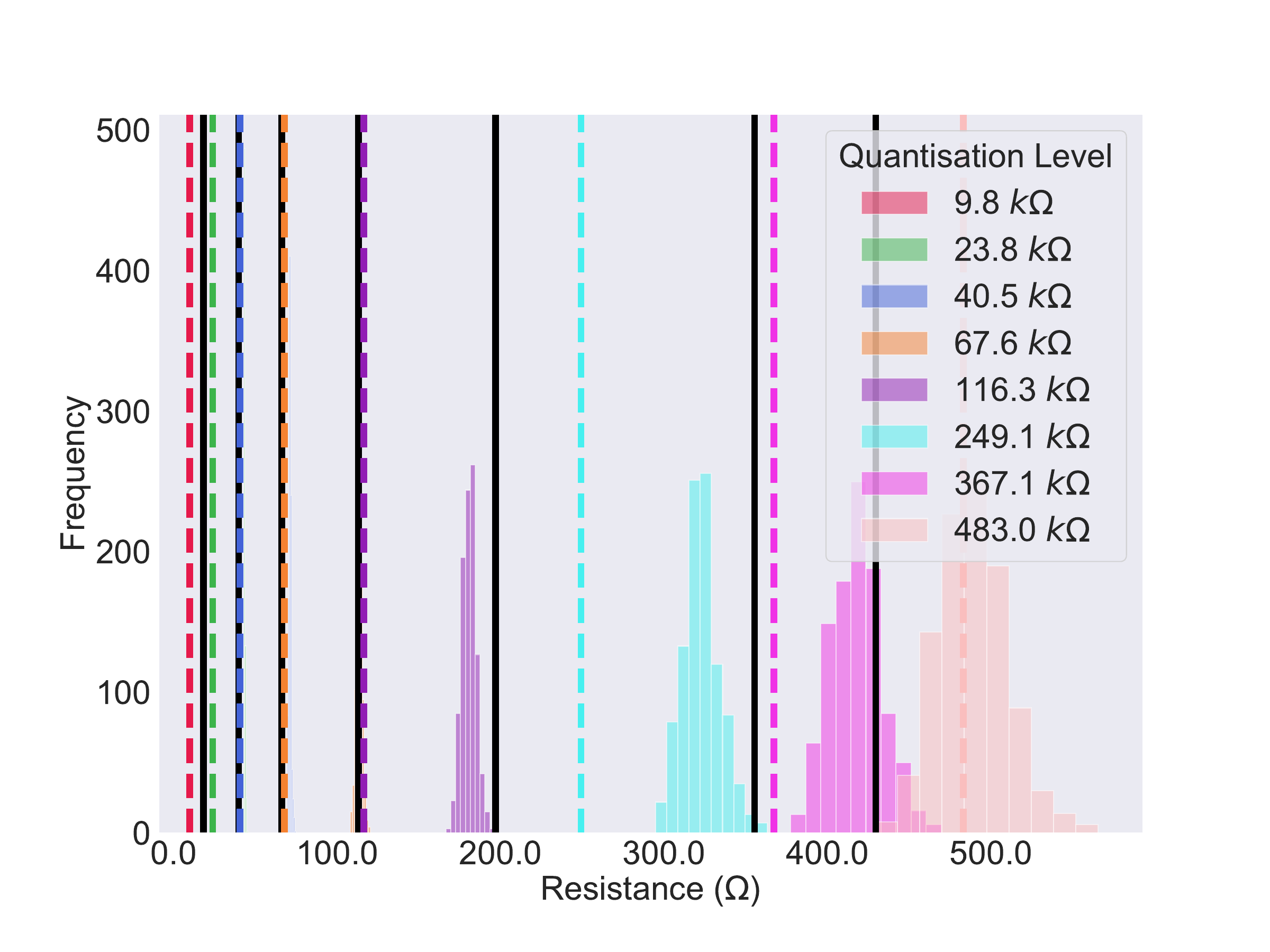}
    \label{fig:quantisation_8}}
    \hspace*{\fill}
    \caption{The end-to-end optimised quantisation schemes learned through using the \ac{cGAN} model as a model of the resistive drift. Figures~\ref{fig:quantisation_2},~\ref{fig:quantisation_4},~and~\ref{fig:quantisation_8} show the schemes trained for a delay of 100 and 2, 4, and 8 quantisation levels (1, 2, 3 bits), respectively.}
    \label{fig:quantisation_plots}
    \Description[Learned quantisation scheme plots.]{Learned optimised quantisation schemes for different numbers of quantisation bits, 2, 4, and 8 bits. All optimised for a delay of 100.}
\end{figure}

Figure~\ref{fig:quantisation_plots} shows the best achieved quantisation scheme for three examples with different delays. We see that for increasing delay, the maximum value of \(|\mathcal{L}|\), and thus the number of bits that can be stored, goes down. We can see that the optimisation procedure tends to result in quantisation levels at the lower end of the learned bins. This corresponds to a correction for the predicted resistive drift in the positive direction (towards the equilibrium resistance of 500\(k\Omega\).

\begin{figure}
    \centering
    \includegraphics[width=.6\linewidth]{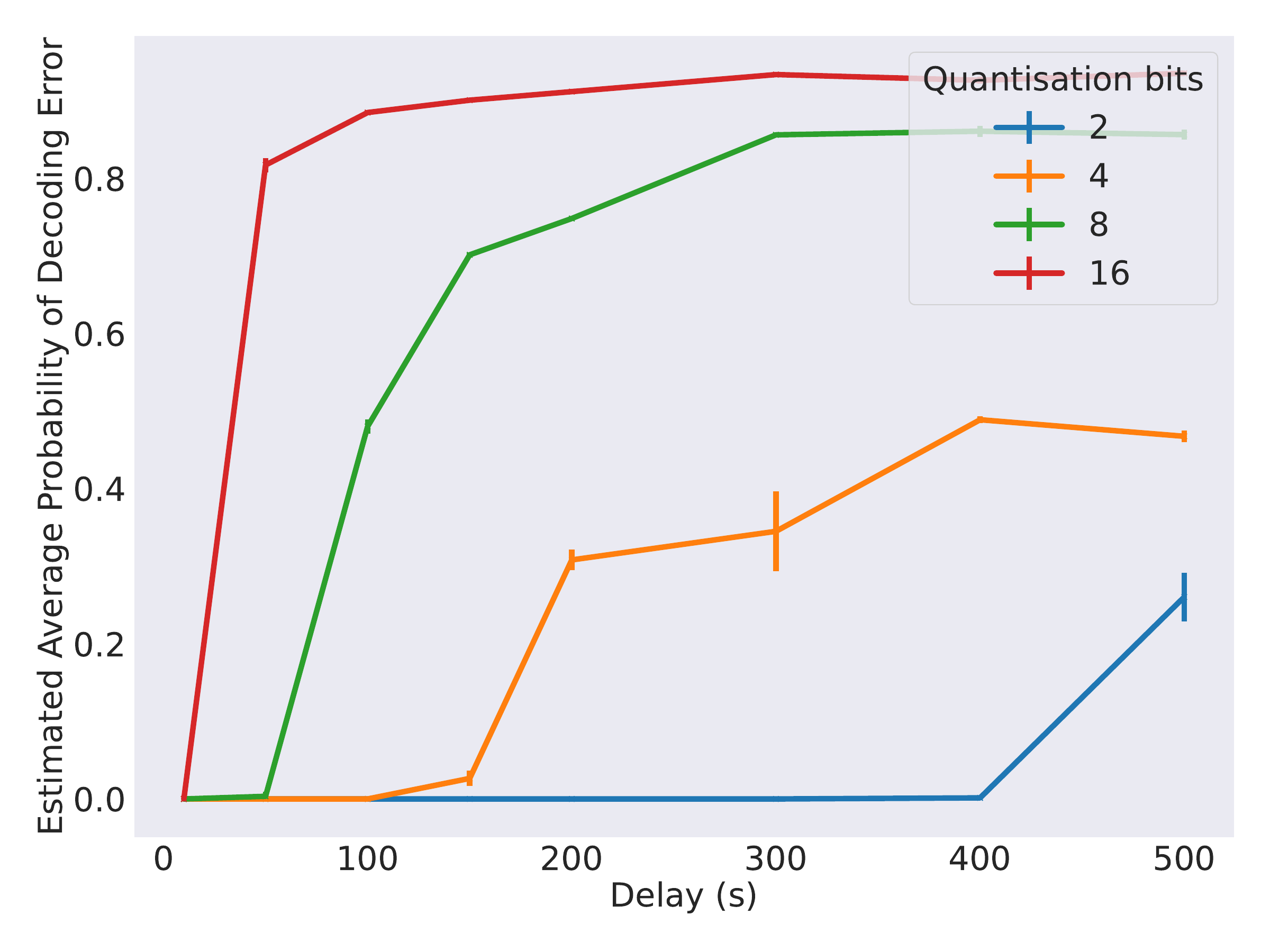}
    \caption{Error vs. delay for different levels of quantisation. As the delay increases, the capacity of the device for storing information decreases, and the decoding error of the optimised quantisation scheme increases, up to a maximum of \((2^b-1)/2^b\), where \(b\) denotes the number of bits of the quantisation scheme. We show results averaged over 5 experiments for each setting.}
    \label{fig:quantisation_all}
    \Description[Quantisation error plot for different numbers of bits and delays.]{The mean decoding error for 2, 4, 8, and 16 quantisation bits, for a range of delays. We see the error increasing with increasing delay.}
\end{figure}

Figure~\ref{fig:quantisation_all} shows the error probability of the optimised levels and decoding boundaries for varying numbers of levels, under different delay conditions. We observe that as the number of quantisation levels increases, so does the error rate. As the delay increases, the delay-conditioned capacity of the memristor for information storage goes down, and it becomes more difficult to achieve quantised storage with a sufficiently low error probability. We see that the error probability approaches \((|\mathcal{L}|-1)/|\mathcal{L}|\) as the delay becomes significantly large, indicating a total loss of information, with the boundary decoding scheme having no advantage over a random guess.

Successful optimisation of the quantisation levels and decoding boundaries for a range of delay conditions highlights the efficacy of our model in providing a computationally simple and differentiable model of the resistive drift conditional distribution, enabling efficient Monte Carlo simulation of the decoding error probabilities in an end-to-end training setup.

\section{Discussion and Conclusion}

In an effort to create a \ac{GAN} that is conditioned directly on the delay condition, we are able to use the technique of delay discrimination to boost the performance and generalisation of the \ac{cGAN} when evaluating at larger delays, improving consistency across the timescales of generation, and enabling the use of the model for a variety of delay conditions.

We note that an alternative viewpoint treats delay discrimination as a form of generative data augmentation, where we assume that the model is more accurate at generating sequences with delays within the observed data, than for those delays without. We use the model's own output sequences for smaller delay values, evaluated recurrently, to learn to generate generate sequences for delay conditions that are much larger than those that would be possible to generate from the finite-length sequences in the dataset.

The \ac{cGAN} generative model presented is completely memoryless, in the sense that we are assuming that given the current voltage input and the state variable (the resistance), knowledge of previous states of the system would not give any benefit. This may not always be the case however, for example, immediately after a programming pulse, when the volatility may initially be high, but may relax over time, meaning there is dependence on past state or input values \cite{el-geresyEventBasedSimulationStochastic2024}. It may therefore be desirable to create a model that is able to encapsulate the history of the device state over time and a recurrent or autoregressive network modelling approach may enable the modelling of this behaviour. An approach to doing this in a computationally simple manner is an left as future work. Future work will also include the evaluation of the delay discriminator method more generally, to assess its ability to improve the training performance of generative time series models at a more fundamental level. The modelling approach as presented could also be extended to alternative generative models and architectures, for example, diffusion models \cite{hoDenoisingDiffusionProbabilistic2020}.

Finally, future work could include adapting the modelling approach for a neuromorphic setting, making use the proposed approach for simulations of neuromorphic circuits. In addition, there exists the opportunity to extend the work to an asynchronous setting, for example through the use of spiking neural networks.

Overall, we have demonstrated that it is possible to effectively model the resistive drift phenomenon in memristors through the use of a \ac{cGAN} trained to model the delay and resistance-conditioned distribution. We saw that we could attempt to tackle multiple challenges associated with the training of adversarial networks for time series data, without the use of supervised losses or scheduled sampling, by making use of a delay discriminator to enforce consistency across a range of timescales. This technique was also shown to extend the scope of the \ac{cGAN}, for generation of delay-conditioned outputs using a single forward pass, to much larger delays than those present in the dataset.

\bibliographystyle{apalike}
\bibliography{references}

\begin{thebibliography}{}

\bibitem[Abbey et~al., 2022]{abbeyThermalEffectsInitial2022}
Abbey, T., Giotis, C., Serb, A., Stathopoulos, S., and Prodromakis, T. (2022).
\newblock Thermal {{Effects}} on {{Initial Volatile Response}} and {{Relaxation Dynamics}} of {{Resistive RAM Devices}}.
\newblock {\em IEEE Electron Device Letters}, 43(3):386--389.

\bibitem[Arjovsky and Bottou, 2016]{arjovskyPrincipledMethodsTraining2016}
Arjovsky, M. and Bottou, L. (2016).
\newblock Towards {{Principled Methods}} for {{Training Generative Adversarial Networks}}.
\newblock In {\em International {{Conference}} on {{Learning Representations}}}.

\bibitem[Arjovsky et~al., 2017]{arjovskyWassersteinGAN2017}
Arjovsky, M., Chintala, S., and Bottou, L. (2017).
\newblock Wasserstein {{GAN}}.
\newblock In {\em {{PMLR}}}, pages 214--223.

\bibitem[Bengio et~al., 2015]{bengioScheduledSamplingSequence2015}
Bengio, S., Vinyals, O., Jaitly, N., and Shazeer, N. (2015).
\newblock Scheduled {{Sampling}} for {{Sequence Prediction}} with {{Recurrent Neural Networks}}.
\newblock In {\em Advances in {{Neural Information Processing Systems}}}, volume~28. Curran Associates, Inc.

\bibitem[Berdan et~al., 2014]{berdanMemristorSPICEModel2014}
Berdan, R., Lim, C., Khiat, A., Papavassiliou, C., and Prodromakis, T. (2014).
\newblock A {{Memristor SPICE Model Accounting}} for {{Volatile Characteristics}} of {{Practical ReRAM}}.
\newblock {\em IEEE Electron Device Letters}, 35(1):135--137.

\bibitem[Carbajal et~al., 2015]{carbajalMemristorModelsMachine2015}
Carbajal, J.~P., Dambre, J., Hermans, M., and Schrauwen, B. (2015).
\newblock Memristor {{Models}} for {{Machine Learning}}.
\newblock {\em Neural Computation}, 27(3):725--747.

\bibitem[Cho et~al., 2014]{choLearningPhraseRepresentations2014}
Cho, K., {van Merri{\"e}nboer}, B., Gulcehre, C., Bahdanau, D., Bougares, F., Schwenk, H., and Bengio, Y. (2014).
\newblock Learning {{Phrase Representations}} using {{RNN Encoder}}--{{Decoder}} for {{Statistical Machine Translation}}.
\newblock In Moschitti, A., Pang, B., and Daelemans, W., editors, {\em Proceedings of the 2014 {{Conference}} on {{Empirical Methods}} in {{Natural Language Processing}} ({{EMNLP}})}, pages 1724--1734, Doha, Qatar. Association for Computational Linguistics.

\bibitem[Chua, 1971]{chuaMemristorTheMissingCircuit1971}
Chua, L. (1971).
\newblock Memristor-{{The}} missing circuit element.
\newblock {\em IEEE Transactions on Circuit Theory}, 18(5):507--519.

\bibitem[Chua, 2011]{chuaResistanceSwitchingMemories2011}
Chua, L. (2011).
\newblock Resistance switching memories are memristors.
\newblock {\em Applied Physics A}, 102(4):765--783.

\bibitem[Chua, 2014]{chuaIfItPinched2014}
Chua, L. (2014).
\newblock If it's pinched it's a memristor.
\newblock {\em Semiconductor Science and Technology}, 29(10):104001.

\bibitem[Chua and {Sung Mo Kang}, 1976]{chuaMemristiveDevicesSystems1976}
Chua, L. and {Sung Mo Kang} (1976).
\newblock Memristive devices and systems.
\newblock {\em Proceedings of the IEEE}, 64(2):209--223.

\bibitem[Ding et~al., 2022]{dingCcGANContinuousConditional2022}
Ding, X., Wang, Y., Xu, Z., Welch, W.~J., and Wang, Z.~J. (2022).
\newblock {{CcGAN}}: {{Continuous Conditional Generative Adversarial Networks}} for {{Image Generation}}.
\newblock In {\em International {{Conference}} on {{Learning Representations}}}.

\bibitem[{El-Geresy} et~al., 2024]{el-geresyEventBasedSimulationStochastic2024}
{El-Geresy}, W., Papavassiliou, C., and G{\"u}nd{\"u}z, D. (2024).
\newblock Event-{{Based Simulation}} of {{Stochastic Memristive Devices}} for {{Neuromorphic Computing}}.
\newblock {\em arXiv:2407.04718 [physics]}.

\bibitem[Goodfellow et~al., 2014]{goodfellowGenerativeAdversarialNetworks2014}
Goodfellow, I.~J., {Pouget-Abadie}, J., Mirza, M., Xu, B., {Warde-Farley}, D., Ozair, S., Courville, A., and Bengio, Y. (2014).
\newblock Generative {{Adversarial Networks}}.
\newblock {\em Advances in neural information processing systems}.

\bibitem[He et~al., 2016]{heDeepResidualLearning2016}
He, K., Zhang, X., Ren, S., and Sun, J. (2016).
\newblock Deep {{Residual Learning}} for {{Image Recognition}}.
\newblock In {\em 2016 {{IEEE Conference}} on {{Computer Vision}} and {{Pattern Recognition}} ({{CVPR}})}, pages 770--778.

\bibitem[Ho et~al., 2020]{hoDenoisingDiffusionProbabilistic2020}
Ho, J., Jain, A., and Abbeel, P. (2020).
\newblock Denoising {{Diffusion Probabilistic Models}}.
\newblock In {\em Advances in {{Neural Information Processing Systems}}}, volume~33, pages 6840--6851. Curran Associates, Inc.

\bibitem[Hochreiter and Schmidhuber, 1997]{hochreiterLongShortTermMemory1997}
Hochreiter, S. and Schmidhuber, J. (1997).
\newblock Long {{Short-Term Memory}}.
\newblock {\em Neural Computation}, 9(8):1735--1780.

\bibitem[Husz{\'a}r, 2015]{huszarHowNotTrain2015}
Husz{\'a}r, F. (2015).
\newblock How (not) to {{Train}} your {{Generative Model}}: {{Scheduled Sampling}}, {{Likelihood}}, {{Adversary}}?
\newblock {\em arXiv:1511.05101 [cs, math, stat]}.

\bibitem[Ielmini, 2011]{ielminiModelingUniversalSet2011}
Ielmini, D. (2011).
\newblock Modeling the {{Universal Set}}/{{Reset Characteristics}} of {{Bipolar RRAM}} by {{Field-}} and {{Temperature-Driven Filament Growth}}.
\newblock {\em IEEE Transactions on Electron Devices}, 58(12):4309--4317.

\bibitem[Kingma and Ba, 2017]{kingmaAdamMethodStochastic2017}
Kingma, D.~P. and Ba, J. (2017).
\newblock Adam: {{A Method}} for {{Stochastic Optimization}}.
\newblock {\em arXiv:1412.6980 [cs]}.

\bibitem[Lin et~al., 2020]{linPacGANPowerTwo2020}
Lin, Z., Khetan, A., Fanti, G., and Oh, S. (2020).
\newblock {{PacGAN}}: {{The Power}} of {{Two Samples}} in {{Generative Adversarial Networks}}.
\newblock {\em IEEE Journal on Selected Areas in Information Theory}, 1(1):324--335.

\bibitem[Malik et~al., 2022]{malikAbsorbingMarkovChain2022}
Malik, A., Papavassiliou, C., and Stathopoulos, S. (2022).
\newblock An {{Absorbing Markov Chain Model}} for {{Stochastic Memristive Devices}}.
\newblock In {\em 2022 11th {{International Conference}} on {{Modern Circuits}} and {{Systems Technologies}} ({{MOCAST}})}, pages 1--4.

\bibitem[Mead, 1990]{meadNeuromorphicElectronicSystems1990}
Mead, C. (1990).
\newblock Neuromorphic electronic systems.
\newblock {\em Proceedings of the IEEE}, 78(10):1629--1636.

\bibitem[Mirza and Osindero, 2014]{mirzaConditionalGenerativeAdversarial2014}
Mirza, M. and Osindero, S. (2014).
\newblock Conditional {{Generative Adversarial Nets}}.
\newblock {\em arXiv:1411.1784 [cs, stat]}.

\bibitem[Molter and Nugent, 2016]{molterGeneralizedMetastableSwitch2016}
Molter, T.~W. and Nugent, M.~A. (2016).
\newblock The {{Generalized Metastable Switch Memristor Model}}.
\newblock In {\em {{CNNA}} 2016; 15th {{International Workshop}} on {{Cellular Nanoscale Networks}} and Their {{Applications}}}, pages 1--2.

\bibitem[Oh et~al., 2019]{ohImpactResistanceDrift2019}
Oh, S., Huang, Z., Shi, Y., and Kuzum, D. (2019).
\newblock The {{Impact}} of {{Resistance Drift}} of {{Phase Change Memory}} ({{PCM}}) {{Synaptic Devices}} on {{Artificial Neural Network Performance}}.
\newblock {\em IEEE Electron Device Letters}, 40(8):1325--1328.

\bibitem[Pickett et~al., 2009]{pickettSwitchingDynamicsTitanium2009}
Pickett, M.~D., Strukov, D.~B., Borghetti, J.~L., Yang, J.~J., Snider, G.~S., Stewart, D.~R., and Williams, R.~S. (2009).
\newblock Switching dynamics in titanium dioxide memristive devices.
\newblock {\em Journal of Applied Physics}, 106(7):074508.

\bibitem[Ronneberger et~al., 2015]{ronnebergerUNetConvolutionalNetworks2015}
Ronneberger, O., Fischer, P., and Brox, T. (2015).
\newblock U-{{Net}}: {{Convolutional Networks}} for {{Biomedical Image Segmentation}}.
\newblock In Navab, N., Hornegger, J., Wells, W.~M., and Frangi, A.~F., editors, {\em Medical {{Image Computing}} and {{Computer-Assisted Intervention}} -- {{MICCAI}} 2015}, Lecture {{Notes}} in {{Computer Science}}, pages 234--241, Cham. Springer International Publishing.

\bibitem[Rumsey, 2019]{rumseyCapacityConsiderationsData2019}
Rumsey, S. (2019).
\newblock {\em Capacity {{Considerations}} for {{Data Storage}} in {{Memristor Arrays}}}.
\newblock PhD thesis, University of Toronto, Toronto, Canada.

\bibitem[Salimans et~al., 2016]{salimansImprovedTechniquesTraining2016}
Salimans, T., Goodfellow, I., Zaremba, W., Cheung, V., Radford, A., and Chen, X. (2016).
\newblock Improved {{Techniques}} for {{Training GANs}}.
\newblock {\em arXiv:1606.03498 [cs]}.

\bibitem[Strukov et~al., 2008]{strukovMissingMemristorFound2008}
Strukov, D.~B., Snider, G.~S., Stewart, D.~R., and Williams, R.~S. (2008).
\newblock The missing memristor found.
\newblock {\em Nature}, 453(7191):80--83.

\bibitem[Sung et~al., 2018]{sungPerspectiveReviewMemristive2018}
Sung, C., Hwang, H., and Yoo, I.~K. (2018).
\newblock Perspective: {{A}} review on memristive hardware for neuromorphic computation.
\newblock {\em Journal of Applied Physics}, 124(15):151903.

\bibitem[{van den Oord} et~al., 2016]{vandenoordWaveNetGenerativeModel2016}
{van den Oord}, A., Dieleman, S., Zen, H., Simonyan, K., Vinyals, O., Graves, A., Kalchbrenner, N., Senior, A., and Kavukcuoglu, K. (2016).
\newblock {{WaveNet}}: {{A Generative Model}} for {{Raw Audio}}.
\newblock {\em arXiv:1609.03499 [cs]}.

\bibitem[Vaswani et~al., 2017]{vaswaniAttentionAllYou2017}
Vaswani, A., Shazeer, N., Parmar, N., Uszkoreit, J., Jones, L., Gomez, A.~N., Kaiser, {\L}., and Polosukhin, I. (2017).
\newblock Attention is all you need.
\newblock In {\em Proceedings of the 31st {{International Conference}} on {{Neural Information Processing Systems}}}, {{NIPS}}'17, pages 6000--6010, Red Hook, NY, USA. Curran Associates Inc.

\bibitem[Williams et~al., 2013]{williamsPhysicsbasedMemristorModels2013}
Williams, R.~S., Pickett, M.~D., and Strachan, J.~P. (2013).
\newblock Physics-based memristor models.
\newblock In {\em 2013 {{IEEE International Symposium}} on {{Circuits}} and {{Systems}} ({{ISCAS}})}, pages 217--220.

\bibitem[Yang et~al., 2008]{yangMemristiveSwitchingMechanism2008}
Yang, J.~J., Pickett, M.~D., Li, X., Ohlberg, D. A.~A., Stewart, D.~R., and Williams, R.~S. (2008).
\newblock Memristive switching mechanism for metal/oxide/metal nanodevices.
\newblock {\em Nature Nanotechnology}, 3(7):429--433.

\bibitem[Yoon and Jarrett, 2019]{yoonTimeseriesGenerativeAdversarial2019}
Yoon, J. and Jarrett, D. (2019).
\newblock Time-series {{Generative Adversarial Networks}}.
\newblock {\em Advances in Neural Information Processing Systems}, pages 5508--5518.

\bibitem[Zarcone et~al., 2020]{zarconeAnalogCodingEmerging2020}
Zarcone, R.~V., Engel, J.~H., Burc~Eryilmaz, S., Wan, W., Kim, S., BrightSky, M., Lam, C., Lung, H.-L., Olshausen, B.~A., and Philip~Wong, H.-S. (2020).
\newblock Analog {{Coding}} in {{Emerging Memory Systems}}.
\newblock {\em Scientific Reports}, 10(1):6831.

\end{thebibliography}

\end{document}